\documentclass[acmsmall,screen]{acmart}
\AtBeginDocument{%
  \providecommand\BibTeX{{%
    Bib\TeX}}}

\usepackage[]{collab}
\usepackage{xcolor}
\usepackage{colortbl}
\usepackage{tcolorbox}
\usepackage{mdframed}
\usepackage{listings}
\usepackage{makecell}
\usepackage{multirow}
\usepackage{array}

\usepackage{amsmath,amssymb,amsfonts}
\usepackage{algorithmic}
\usepackage{graphicx}
\usepackage{fbox}
\usepackage{textcomp}
\usepackage{minibox}
\usepackage{balance}
\usepackage{enumitem}
\usepackage{xspace}
\usepackage[T1]{fontenc}
\usepackage{booktabs}
\usepackage{threeparttable}
\usepackage{url}
\usepackage{soul}
\usepackage{ulem}
\usepackage{balance}
\usepackage{subfigure}
\usepackage{diagbox}

\def\BibTeX{{\rm B\kern-.05em{\sc i\kern-.025em b}\kern-.08em
    T\kern-.1667em\lower.7ex\hbox{E}\kern-.125emX}}

\newcommand\num{705\xspace}

\newcommand{\ie}{{\em i.e.},\xspace}
\newcommand{\eg}{{\em e.g.},\xspace}

\definecolor{ballblue}{rgb}{0.13, 0.67, 0.8}
\definecolor{grey}{rgb}{0.9, 0.9, 0.9}
\definecolor{googlered}{rgb}{0.914, 0.262, 0.207}
\definecolor{dandelion}{rgb}{0.95, 0.65, 0.0}
\definecolor{citecolor}{RGB}{106, 34, 107}

\definecolor{ballblue}{rgb}{0.13, 0.67, 0.8}
\definecolor{jcpink}{RGB}{255, 0, 96}
\collabAuthor{gb}{orange}{Guangba}
\collabAuthor{jc}{jcpink}{JC}
\collabAuthor{zh}{green}{Zhihan Jiang}
\collabAuthor{jj}{purple}{Junjie}
\collabAuthor{zr}{teal}{Zirui}
\collabAuthor{yl}{red}{Yulun}
\collabAuthor{yc}{blue}{Yichen}
\collabAuthor{ry}{dandelion}{Renyi}
\collabAuthor{yd}{orange}{Yuedong}

\definecolor{mygreen}{HTML}{AFCFA5}
\newcounter{summary}

\definecolor{mygreen}{HTML}{AFCFA5}
\newcounter{implication}

\definecolor{myyellow}{HTML}{FFF2CC}
\newcounter{finding}
\newcommand{\finding}[1]{\refstepcounter{finding}
	\begin{mdframed}[linecolor=gray!25,roundcorner=12pt,backgroundcolor=mygreen!30,linewidth=3pt,innerleftmargin=2pt, leftmargin=0cm,rightmargin=0cm,topline=false,bottomline=false,rightline=false,leftline=false,skipabove=3pt]
		\textbf{Finding \arabic{finding}:} #1
	\end{mdframed}
}

\setcopyright{acmlicensed}
\copyrightyear{2026}
\acmYear{2026}
\acmDOI{XXXXXXX.XXXXXXX}
\acmConference[Conference acronym 'XX]{Make sure to enter the correct
  conference title from your rights confirmation email}{June 03--05,
  2018}{Woodstock, NY}
\acmISBN{978-1-4503-XXXX-X/2018/06}

\begin{document}

\title{Why Does the LLM Stop Computing: An Empirical Study of User-Reported Failures in Open-Source LLMs}


\author{Guangba Yu}
\orcid{0000-0001-6195-9088}
\authornote{Equal Contribution.}
\email{guangbayu@cuhk.edu.hk}
\affiliation{%
  \institution{The Chinese University of HongKong}
  \city{Hong Kong SAR}
  \country{China}
}

\author{Zirui Wang}
\authornotemark[1]
\email{wangzr39@mail2.sysu.edu.cn}
\orcid{0009-0004-5773-8716}
\affiliation{%
  \institution{Sun Yat-sen University}
  \city{Zhuhai City}
  \country{China}
}

\author{Yujie Huang}
\email{yjhuang@cse.cuhk.edu.hk}
\affiliation{%
  \institution{The Chinese University of HongKong}
  \city{Hong Kong SAR}
  \country{China}
}

\author{Renyi Zhong}
\email{ryzhong22@cse.cuhk.edu.hk}
\affiliation{%
  \institution{The Chinese University of HongKong}
  \city{Hong Kong SAR}
  \country{China}
}

\author{Yuedong Zhong}
\email{zhongyd6@mail2.sysu.edu.cn}
\affiliation{%
  \institution{The Chinese University of HongKong}
  \city{Hong Kong SAR}
  \country{China}
}

\author{Yilun Wang}
\email{ylwang25@cse.cuhk.edu.hk}
\affiliation{%
  \institution{The Chinese University of HongKong}
  \city{Hong Kong SAR}
  \country{China}
}


\author{Michael R. Lyu}
\orcid{0000-0002-3666-5798}
\email{lyu@cse.cuhk.edu.hk}
\affiliation{%
  \institution{The Chinese University of HongKong}
  \city{Hong Kong SAR}
  \country{China}
}

\renewcommand{\shortauthors}{Guangba et al.}

\begin{abstract}

The democratization of open-source Large Language Models (LLMs) allows users to fine-tune and deploy models on local infrastructure but exposes them to a ``First Mile'' deployment landscape. Unlike black-box API consumption, the reliability of user-managed orchestration remains a critical blind spot. To bridge this gap, we conduct the first large-scale empirical study of 705 real-world failures from the  open-source DeepSeek, Llama, and Qwen ecosystems.

Our analysis reveals a paradigm shift: white-box orchestration relocates the reliability bottleneck from model algorithmic defects to the systemic fragility of the deployment stack. We identify three key phenomena: (1) Diagnostic Divergence: runtime crashes distinctively signal infrastructure friction, whereas incorrect functionality serves as a signature for internal tokenizer defects. (2) Systemic Homogeneity: Root causes converge across divergent series, confirming reliability barriers are inherent to the shared ecosystem rather than specific architectures. (3) Lifecycle Escalation: Barriers escalate from intrinsic configuration struggles during fine-tuning to compounded environmental incompatibilities during inference. Supported by our publicly available
dataset, these insights provide actionable guidance for enhancing the reliability of the LLM landscape.
\end{abstract}





\maketitle

\section{Introduction}\label{sec:introduction}
Large Language Models (LLMs) have become a foundational technology in artificial intelligence (AI)~\cite{ChatGPT,wang2026ainativebench}, and the proliferation of open-source series like DeepSeek~\cite{DeepSeek}, Llama~\cite{Llama}, and Qwen~\cite{Qwen} is accelerating their widespread adoption. With these openly accessible models, users are increasingly moving beyond API calls to deploy and fine-tune LLMs on their own infrastructure~\cite{SGLang,vllm}. This democratization of advanced AI fosters immense innovation, but it also exposes users to significant new reliability challenges.

The fine-tuning and deployment of open-source LLMs introduce unique failure classes not typically seen in traditional software~\cite{XiaoyunFaults2022ISSRE,HaryadiOutages2016SoCC,Microsoft2022SoCC} or even other deep learning (DL) systems~\cite{DLFailure2019FSE,DLFailure2020ICSE,Junjie2023Tosem,DependencyBug2023FSE,FailureSurvey}. These failures arise from a confluence of factors: (1) the \textbf{massive scale and computational demands}, leading to infrastructure, memory (\eg Key-Value (KV) Cache \cite{vllm}), and distributed computing issues; (2) their \textbf{emergent and non-deterministic behaviors}, which manifest as semanticfailures like hallucinations and instruction-following errors, distinct from traditional DL system failures; and (3) the growing ecosystem of \textbf{specialized fine-tuning and inference methods} (\eg LoRA \cite{Lora1,Lora2}, quantization \cite{Quantization1,Quantization2}, and inference frameworks \cite{SGLang,vllm}), which creates novel bugs related to numerical precision and compatibility.

Despite the rapid adoption of LLMs, our understanding of their reliability remains fragmented. 
Existing research has predominantly focused on either the \textit{internal} stability of large-scale model training and cloud services~\cite{Shangtang2024NSDI,Meta2025HPCA,yan2025empiricalstudyproductionincidents}, or the \textit{external} usage of black-box APIs~\cite{chen2025empiricalstudyopenaiapi}. 
While valuable, these perspectives leave a critical gap: they either analyze failures that are invisible to the end-user (internal service outages) or focus on scenarios where the underlying infrastructure is completely abstracted away (API consumption).

Consequently, the \textit{runtime reliability} of white-box models deployed on user-managed infrastructure remains an unexplored frontier. Unlike the opaque, provider-managed Pre-training phase, our study concentrates exclusively on the two user-facing stages: Fine-tuning (adapting models with domain data) and Inference (serving models for generation).
This ``First Mile'' of deployment, where users must orchestrate the entire stack from CUDA kernels to quantization formats, is fraught with unique dynamic failures (\eg \textit{Quantization Instability}, \textit{KV Cache Corruption}) that are fundamentally different from API-level errors or training divergences.

To bridge this gap, we conduct the first large-scale empirical study of \num real-world LLM failures collected from three prominent open-source LLM series: DeepSeek, Llama, and Qwen. This unique dataset, encompassing multiple models and versions, is collected and manually labeled through a systematic process detailed in \S~\ref{sec:method}. Based on this dataset (details in Table~\ref{tab:model_overview}), our study provides a holistic understanding of LLM failures from the user's perspective, with key findings addressing six research questions (RQs):
\begin{enumerate}[leftmargin=*, label=\textbullet, noitemsep, topsep=2pt]
    \item \textbf{RQ1 (Symptom): What are the symptoms of failures encountered by open-source LLM users?} 
    We identify a ``First Mile'' barrier where \textit{Runtime Crashes} (61.1\%) dominate due to infrastructure friction. Furthermore, we characterize \textit{Generation Quality Anomalies} not as mere bugs, but as a paradigm shift toward probabilistic cognitive deviations (\S~\ref{sec:rq1}).
    
    \item \textbf{RQ2 (Root Cause): What are the underlying root causes of LLM failures?} 
    We reveal the systemic fragility of the software stack. External factors, specifically \textit{Environment Faults} (45.7\%) driven by \textbf{rigid coupling} and \textit{User Configuration Faults} (25.4\%) caused by distinct usability deficits, have largely replaced traditional model-centric coding bugs (\S~\ref{sec:rq2}).
    
    \item \textbf{RQ3 (Symptom-Cause Relation): What is the relationship between observed symptoms and their root causes?} 
    Our mapping establishes distinct \textbf{diagnostic heuristics}: \textit{Runtime Crashes} serve as predominant indicators of infrastructure friction, whereas \textit{Incorrect Functionality} acts as a distinctive signature for specific \textit{Tokenizer and Text Processing Defects} (\S~\ref{sec:rq3}).
    
    \item \textbf{RQ4 (Comparative Analysis): How do LLM failures differ between three model series?} 
    We uncover a systemic homogeneity beneath surface-level differences. While symptom distributions vary locally, the underlying root causes remain identical across series, confirming that reliability challenges are inherent to the shared ecosystem rather than specific model architectures (\S~\ref{sec:rq4}).
    
    \item \textbf{RQ5 (Lifecycle Stage): How do the characteristics of LLM failures differ between the fine-tuning and inference stage?} 
    We define a trajectory of escalation. While intrinsic configuration complexity remains a persistent hurdle, Inference compounds this challenge with a dominant influx of extrinsic operational friction, expanding the scope of reliability barriers (\S~\ref{sec:rq5}).
    
    \item \textbf{RQ6 (Failure Evolution): How have LLM failure patterns evolved with new model releases?} 
    We trace a maturation trajectory: as foundational stability issues are resolved, the failure landscape shifts toward the structural complexities of advanced optimization (e.g., quantization) and the increasing environmental demands of scaling (\S~\ref{sec:rq6}).
\end{enumerate}

In summary, this paper makes the following main contributions:
\begin{itemize}[leftmargin=*,label=\textbullet, noitemsep, topsep=2pt]
    \item \textbf{A New Perspective on the LLM Failure Landscape.} We present the first large-scale empirical study of \num user-reported failures for open-source LLMs. By contrasting these with traditional DL systems, we reveal a fundamental paradigm shift in failure patterns and highlight the unique challenges of the LLM ecosystem (\S~\ref{sec:study}).
    \item \textbf{A Data-Backed Diagnosis and Action Guidance.} Our research provides a comprehensive set of actionable insights for practitioners and providers. We show that the primary reliability challenges in the democratized LLM ecosystem do not stem from intrinsic model flaws, but from the immense complexity of the environmental, configuration, and deployment stacks (\S~\ref{sec:lesson}).
    \item \textbf{A Publicly Available Research Artifact.} To facilitate future research, we have made our manually labeled dataset of \num LLM failures and analysis results publicly available~\cite{artifact}.
\end{itemize}

\begin{table}[t] 
\centering
\caption{Overview of selected open-source LLM series.}
\vspace{-0.15in}
\label{tab:model_overview}
\resizebox{0.8\columnwidth}{!}{%
\begin{tabular}{@{}llccc@{}}
\toprule
\textbf{Series} & \textbf{Model} & \textbf{\#Star} & \textbf{\#All Issues} & \textbf{\#Selected Issues} \\ \midrule
DeepSeek~\cite{DeepSeek} & Coder, Coder-V2, Math, MoE, R1, V3       & 223.2k &1475  &86  \\
LLaMA~\cite{Llama}           & CodeLlama, Llama 2, Llama 3& 103.3k &1578 &170                     \\
Qwen~\cite{Qwen}     & Qwen, Qwen2.5-Coder, Qwen2.5, Qwen2.5-Math & 45.7k  &2249  &449  \\ \midrule
Total & &372.2k &5302 & \num \\ \bottomrule
\end{tabular}%
}\\[0.02in] %
{\footnotesize \parbox{0.75\columnwidth}{\raggedright Note: Data collected before 23 April 2025.}}
\vspace{-0.1in}
\end{table}

\section{Empirical Study Methodology}\label{sec:method}

\begin{figure*}[t]
    \centering
    \includegraphics[width=0.9\textwidth]{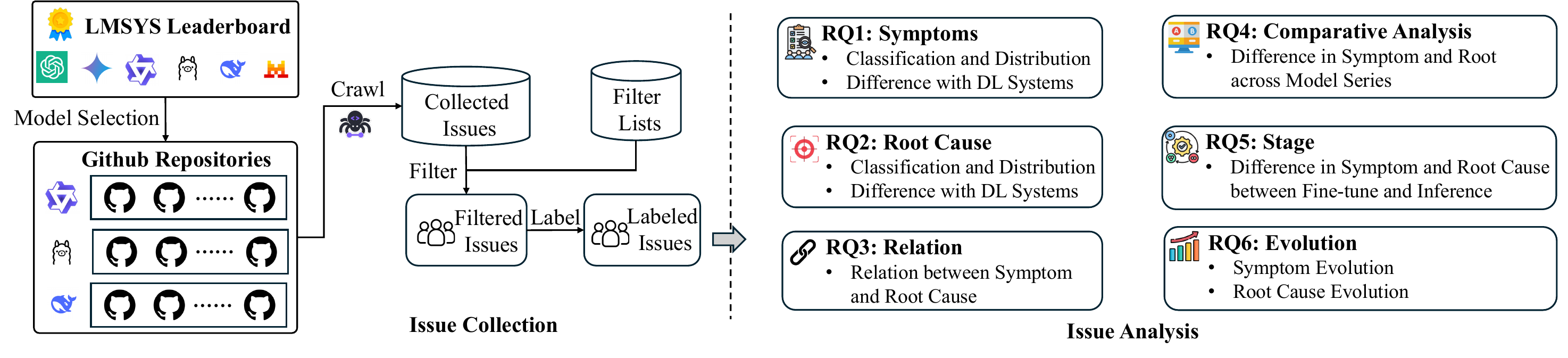}
    \vspace{-0.15in}
    \caption{Overview of empirical study methodology.}
    \label{fig:method}
    \vspace{-0.2in}
\end{figure*}

Figure~\ref{fig:method} outlines our methodology. Given the nascent nature of the open-source LLM ecosystem, standard automated mining often struggles with the novelty of failure modes. Therefore, we designed a \textbf{Human-Centric Analysis} pipeline, prioritizing the fidelity of diagnosis over the quantity of data to establish a rigorous ``Ground Truth.''

\subsection{Targeted Model Selection}~\label{sec:model_selection}
To construct a representative dataset, we designed a three-stage selection funnel:

\textbf{(1) Defining the Scope:} We exclusively selected open-source LLMs that maintain an official, active GitHub repository as the central hub for issue reporting. This ensures our analysis captures the direct feedback loop between end-users and core developers.

\textbf{(2) Surveying and Filtering:} Based on leaderboards LMArena~\cite{LMArena}, we surveyed prominent series including Llama, Qwen, DeepSeek, Mistral, and Gemma. We prioritized repositories with at least 500 stars and consistent issue engagement. Notably, we excluded series like Mistral's primary repository~\cite{mistral-inference} as it contained fewer than 30 closed issues, providing an insufficient sample size for robust empirical analysis.

\textbf{(3) Final Selection:} We selected \textbf{DeepSeek}~\cite{DeepSeek}, \textbf{Llama}~\cite{Llama}, and \textbf{Qwen}~\cite{Qwen}. These series are ideal for two reasons: (1) \textit{Market Representativeness}: They are consistently ranked among leaders in user adoption; (2) \textit{Diversity}: They originate from three distinct technology organizations (DeepSeek, Meta, and Alibaba), mitigating the bias of a single company's engineering culture.

\subsection{Failure-related Issue Collection}
\label{subsec:collection}
Our data collection process began by retrieving the complete history of 5,302 issues from the official GitHub repositories of the selected LLM series. 
To ensure consistency throughout the filtering process, we established a formal definition of an \textit{LLM Failure} for this study: 
\textit{A user-reported deviation from expected behavior that occurs during the fine-tuning or inference stages, encompassing both explicit system faults (\eg crashes, hangs, error codes) and implicit model anomalies (\eg incoherent generation, endless repetition).}
Guided by this definition, we implemented a rigorous three-stage filtering pipeline to distill the raw data into a high-fidelity dataset.

\textbf{Stage 1: Automated Keyword Filtering.} 
To identify potential reports matching our definition, we queried issues containing failure-related terms in their title or body. We utilized a comprehensive search string designed to capture diverse manifestations of reliability issues: 
\texttt{"error", "bug", "crash", "exception", "hang", "freeze", "slow", "fail", "failure", "fault", "invalid", "incorrect", "problem", "unexpected"}. 
This broad-spectrum search reduced the initial pool to 2,176 candidate issues for manual inspection.

\textbf{Stage 2: Diagnostic Manual Review.} 
Automated filtering inevitably introduces noise (\eg feature requests containing ``error handling''). To mitigate this, two researchers independently conducted a diagnostic review of the titles and descriptions of the 2,176 candidates. 
Based on our failure definition, we strictly excluded: (1) Non-Failures: Feature requests, enhancement proposals, and general usage inquiries (\eg ``How to verify download?''). (2) Out-of-Scope Issues: Failures occurring during pre-training (developer-side) rather than user deployment. (3) Duplicates: Explicitly linked duplicate reports.
Adopting a conservative retention strategy, we narrowed the set to 1,180 high-potential candidates.

\textbf{Stage 3: Final In-Depth Validation.} 
In the final stage, the researchers performed a deep-dive analysis of the full context (including logs, screenshots, and maintainer discussions) for the remaining 1,180 candidates. We retained only issues where a definitive failure was confirmed. Disagreements were resolved by a third senior author acting as an arbiter.

This multi-stage process yielded our final dataset of \num verified failure-related issues. The different issue counts in Table~\ref{tab:model_overview} among the three model series do not indicate bias in our methodology. Instead, they reflect the real-world reporting landscape, which is influenced by factors such as community size, model release dates, and popularity. 


\subsection{Data Labeling}\label{sec:data_labeling}
Following data collection, we proceeded to categorize the \num issues. This phase was guided by a dedicated rationale to ensure scientific rigor.

\subsubsection{Rationale for Manual Analysis}
Although LLMs show potential in automating SE tasks, we adopted a rigorous manual analysis approach. This choice aligns with established standards in empirical software engineering~\cite{TensorFlow2018ISSTA,DLFailure2019FSE,DLFailure2020ICSE,Junjie2023Tosem} and is driven by three domain-specific factors:
\begin{enumerate}[leftmargin=*, noitemsep, topsep=2pt]
    \item \textbf{Handling Novel Failure Modes:} Emerging failures (\eg \textit{Quantization Incompatibility} in FP8 kernels) are often absent from LLM training data. Automation risks hallucinating these novel symptoms into generic categories.
    \item \textbf{Contextual Diagnosis Complexity:} Accurate diagnosis often requires synthesizing implicit cues from fragmented logs and multi-turn discussions, which is a complex task where automated tools currently struggle to achieve high fidelity.
    \item \textbf{Establishing a Ground Truth:} As the first comprehensive study of this ecosystem, our objective is to establish a high-quality Ground Truth dataset free from the stochastic biases of LLM-generated labels.
\end{enumerate}

\subsubsection{Taxonomy Development}
We employed an open coding procedure~\cite{opencoding} guided by the MECE principle (Mutually Exclusive and Collectively Exhaustive)~\cite{lee2018mutually} to ensure orthogonality. 
The process involved three stages: (1) Two annotators independently analyzed a random subset (20\%) to generate initial codes; (2) Codes were collaboratively grouped into hierarchical structures; (3) The taxonomy was validated on a new unseen subset until saturation was reached.
This iterative process yielded two fine-grained hierarchical taxonomies: (1) \textbf{Symptom Taxonomy:} Consists of 4 high-level categories decomposing into 33 specific leaf nodes (\eg \textit{Generation Quality Anomaly}), classifying the user-observable state (\S~\ref{sec:rq1}). (2) \textbf{Root Cause Taxonomy:} Consists of 5 high-level categories with 27 leaf nodes (\eg \textit{KV Cache Allocation Error}), identifying the precise technical fault location (\S~\ref{sec:rq2}).
This multi-level structure ensures our analysis captures both macro-level reliability trends and micro-level debugging details.

\subsubsection{Labeling Procedure and Quality Control}
The labeling of the full dataset was governed by a strict quality control protocol to ensure objectivity and reproducibility. The process involved three distinct phases executed by a team of six SE researchers:

\textbf{Phase 1: Team-wide Calibration.} To mitigate subjective bias, we first developed a standardized coding Guide containing precise definitions and boundary examples for all 60 leaf categories. The team conducted pilot labeling on a set of 50 distinct issues, discussing discrepancies iteratively until a shared consensus on the taxonomy's application was reached.
    
\textbf{Phase 2: Independent Dual-Labeling.} Each of the \num issues was randomly assigned to two annotators. Working independently, they analyzed the full issue context to assign labels for three dimensions: Symptom (RQ1), Root Cause (RQ2), and Lifecycle Stage (RQ5).
    
\textbf{Phase 3: Reliability Assessment and Arbitration.} We quantitatively validated the process by calculating Cohen's Kappa coefficient~\cite{kappa}. The inter-annotator agreement (IAA) remained consistently high, exceeding 0.92 across all dimensions. According to previous study~\cite{landis1977measurement}, this score signifies ``almost perfect agreement,'' confirming that our manual analysis is robust and reproducible. Finally, the few conflicting cases were resolved through a centralized arbitration process led by a senior author to ensure the final dataset's integrity.

\section{Empirical Study Results}\label{sec:study}

\begin{figure*}[t]
    \centering
    \includegraphics[width=0.8\textwidth]{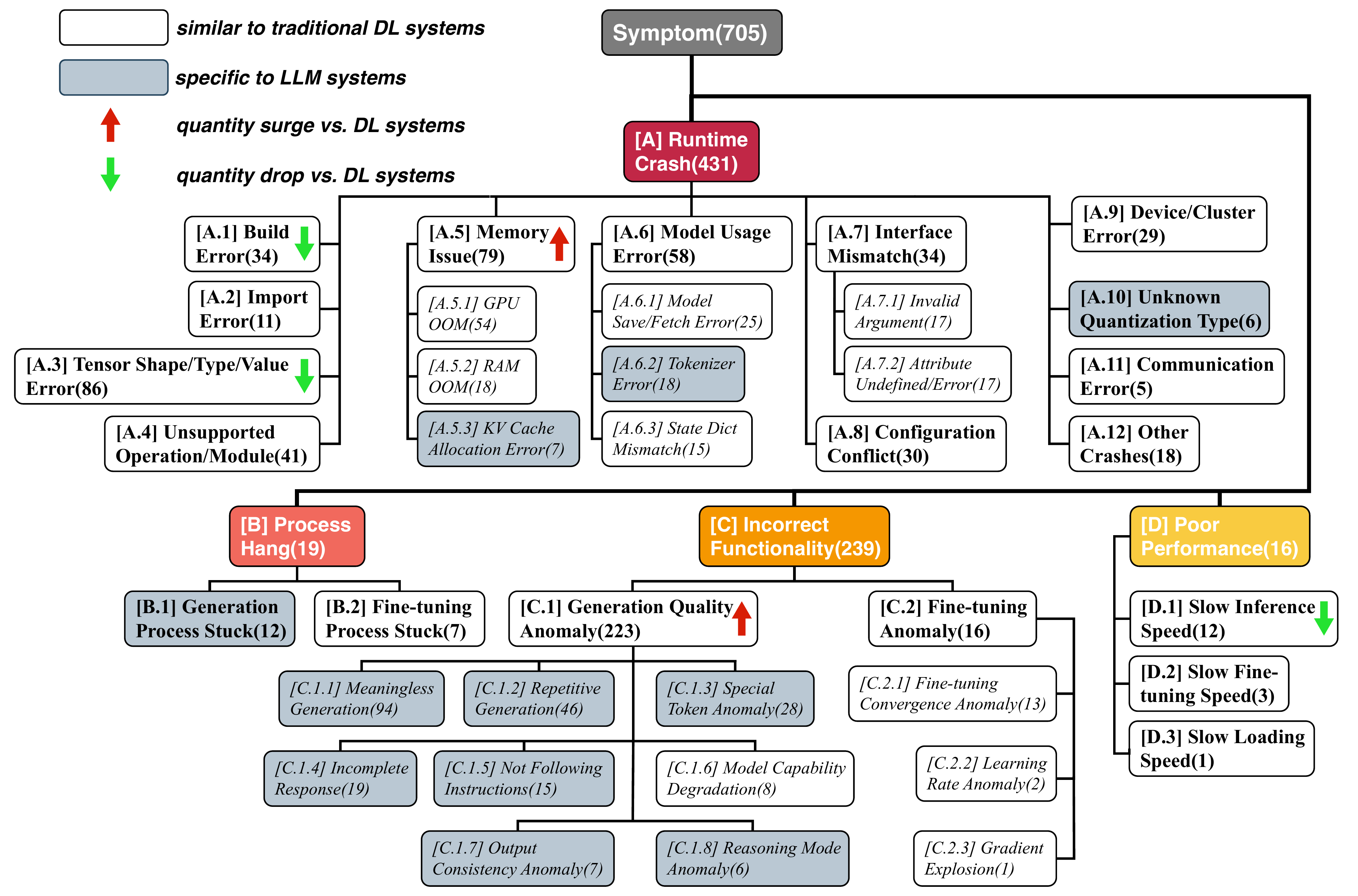}
    \vspace{-0.1in}
    \caption{Symptom taxonomy of failure in open-source LLMs.}
    \label{fig:symptom_taxonomy}
    \vspace{-0.2in}
\end{figure*}

\subsection{RQ1: Symptoms of LLM Failures}\label{sec:rq1}
Figure~\ref{fig:symptom_taxonomy} shows the taxonomy and distribution of symptoms in open-source LLM failures based on the program's observable state from the user's perspective. It is grouped into four high-level categories (\ie Runtime Crash, Process Hang, Incorrect Functionality, and Poor Performance). The gray rectangles denote LLM-specific symptoms (\eg Repetitive Generation) and white rectangles indicate symptoms shared with DL systems (\eg Import Error).

\subsubsection{Runtime Crash [A]}
Our data show that the user's journey is dominated by hard, deterministic failures where the program stops, which we categorize as \textit{Runtime Crash} [A]. This category is the most prevalent by a large margin, accounting for 431 of \num issues (61.1\%). These failures represent a formidable barrier that users need to overcome before they can even evaluate an LLM's capabilities. 

\begin{figure*}[t]
    \centering
    \includegraphics[width=0.95\textwidth]{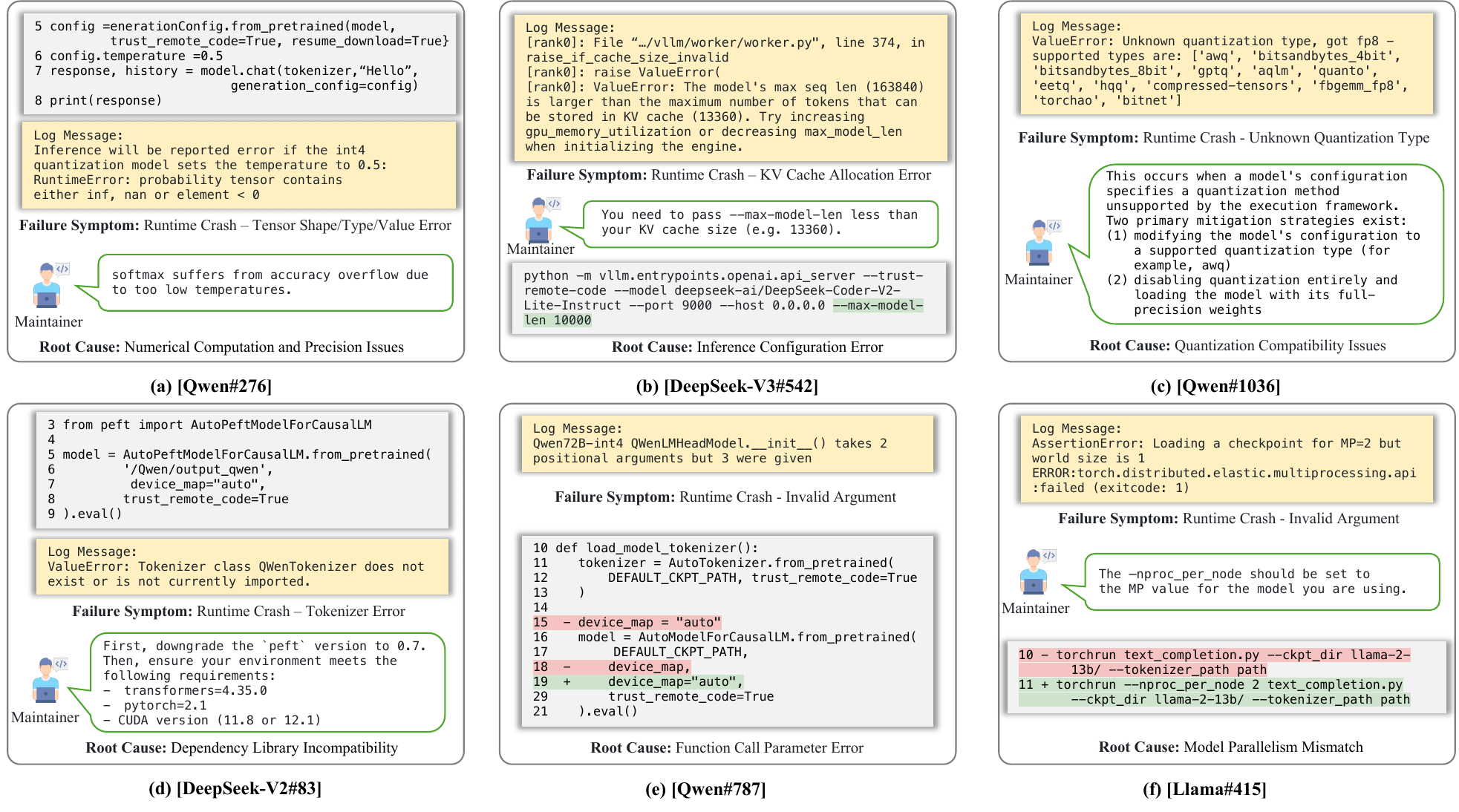}
    \vspace{-0.1in}
    \caption{Illustrative cases of various Runtime Crash failures reported in GitHub issues.}
    \label{fig:case1}
    \vspace{-0.2in}
\end{figure*}

A substantial number of crashes occur during the initial environment setup. Common symptoms include \textit{Build Error} [A.1] (34 instances), which typically manifests as a compilation failure of CUDA kernel \href{https://github.com/meta-llama/llama/issues/104}{[Llama\#104]}, and \textit{Import Error} [A.2] (11 instances). An \textit{Import Error} often signifies a deep dependency version conflict \href{https://github.com/QwenLM/Qwen/issues/460}{[Qwen\#460]}).
The most frequent crash symptom is \textit{Tensor Shape/Type/Value Error} [A.3], with 86 instances. This category encompasses a range of issues in which misformed or incompatible tensor data lead to termination. A frequent manifestation is \texttt{TypeError} resulting from mismatched data types. A notable issue in LLMs where mixed-precision training (\eg \texttt{float32} vs. \texttt{bfloat16} in \href{https://github.com/meta-llama/llama/issues/778}{[Llama\#778]}) can lead to incompatibilities. Another common form is a \texttt{ShapeError} from incompatible tensor dimensions, a critical issue within the attention mechanisms of LLM. In addition, \texttt{InvalidValue} errors occur when data is out of the expected range, such as a probability tensor contains \texttt{inf} or \texttt{nan} elements during generation (\eg Fig.~\ref{fig:case1}(a)).

The sheer scale of modern LLMs exacerbates traditional resource-related failures. 
\textit{Memory Issue} [A.5] (79 instances) is the second leading cause of crashes, headlined by GPU OOM (Out of Memory) [A.5.1] (54 instances) \href{https://github.com/deepseek-ai/DeepSeek-Coder/issues/78}{[DeepSeek-Coder\#78]}. The analysis also reveals the significance of \textit{RAM OOM} [A.5.2] (18 instances), often during the initial model loading phase. A novel, LLM-specific failure in this category is the \textit{KV Cache Allocation Error} [A.5.3] (7 instances), which manifests itself as a crash when the model's maximum sequence length exceeds the KV cache capacity (Fig.~\ref{fig:case1}(b)). Furthermore, the need for distributed deployment in LLMs leads to frequent \textit{Device/Cluster Error} [A.9] (29 instances) \href{https://github.com/deepseek-ai/DeepSeek-R1/issues/532}{[DeepSeek-R1\#532]}.

Crashes also frequently arise from compatibility issues with the execution framework. \textit{Unsupported Operation/Module} [A.4] (41 instances) occurs when a model feature is not supported by the user's environment, leading to errors like \texttt{Target module QuantLinear() is not supported} \href{https://github.com/QwenLM/Qwen/issues/431}{[Qwen\#431]}. The widespread need for quantization, driven by memory constraints, also introduces failures like \textit{Unknown Quantization Type} [A.10] (6 instances),  where a crash is caused by using an unsupported \texttt{fp8} quantization type (Fig.~\ref{fig:case1}(c)).

A substantial number of crashes are caused by incorrect user interaction and complex deployment configurations. \textit{Model Usage Error} [A.6] (58 instances) includes failures in loading models and crashes within the tokenizer itself. For example, as shown in Fig.~\ref{fig:case1}(d), an error in \texttt{Tokenizer} can also cause the system to crash. Closely related is \textit{Interface Mismatch} [A.7] (34 instances), where users provide invalid arguments or encounter undefined attributes when calling the API. Fig.~\ref{fig:case1}(e) shows an \textit{Invalid Argument} example, where a model's initialization method received an incorrect number of positional arguments. Furthermore, \textit{Configuration Conflict} [A.8] (30 instances) arises from invalid settings in user configuration files. Fig.~\ref{fig:case1}(f) shows a classic example, where a crash is triggered by a configuration conflict between a multi-GPU checkpoint (\texttt{MP=2}) and a single-GPU runtime environment (\texttt{world size is 1}).

\finding{The ecosystem is bottlenecked by basic runtime stability (61.1\% crashes) rather than model capability. This infrastructure friction, driven primarily by dependency conflicts and resource exhaustion, blocks users from even accessing the model's fundamental utility.}\label{fnd:runtime_crash}

\begin{figure*}[t]
    \centering
    \includegraphics[width=0.95\textwidth]{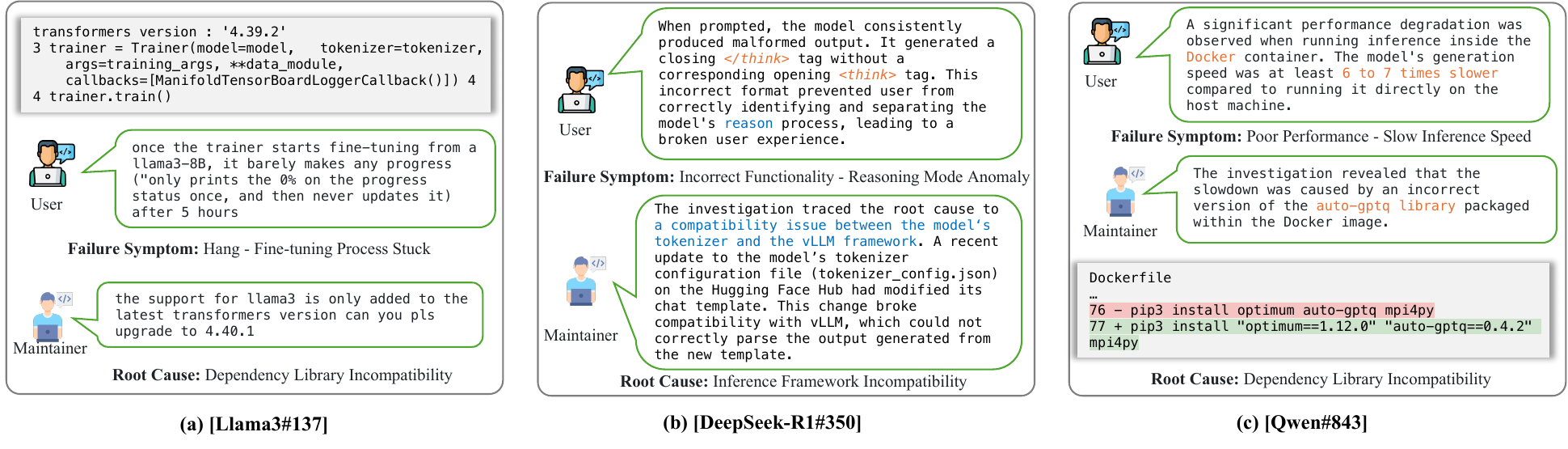}
    \vspace{-0.1in}
    \caption{Illustrative cases of non-crashing failures reported in GitHub issues.}
    \label{fig:case2}
    \vspace{-0.2in}
\end{figure*}

\subsubsection{Process Hang [B]}
A less frequent but disruptive symptom is the \textit{Process Hang}, where an LLM process becomes completely unresponsive without crashing. This category accounts for 19 issues (2.7\%).  The primary manifestations are \textit{Generation Process Stuck} [B.1] (12 instances), where the model freezes during inference, and \textit{Fine-tuning Process Stuck} [B.2] (7 instances), where the fine-tuning process stalls, preventing further progress in model adaptation or training iterations. 
A typical example is shown in Fig.~\ref{fig:case2}(a), where a user reported that a fine-tuning job on a Llama 3 model made no progress for over five hours, with the progress bar stuck at 0\%. This illustrates a silent failure where the process becomes completely unresponsive without an error message.

\subsubsection{Incorrect Functionality [C]}
This category captures the core semantic challenges unique to Generative AI. Accounting for 239 issues (33.9\%), these failures differ fundamentally from traditional software bugs: they manifest not as program terminations, but as probabilistic deviations in cognitive behavior. Unlike crashes which block execution, these anomalies degrade the utility of the model, making them more insidious and difficult to detect in automated pipelines.

\textit{Generation Quality Anomaly} [C.1] dominates this category with 225 instances. This symptom impacts the usability of LLM outputs by producing incoherent, irrelevant, or inaccurate text, which is a key challenge for generative models. \textit{Meaningless Generation} [C.1.1] (94 instances) is the most frequent sub-type, where the model generates nonsensical or garbled text (\eg \href{https://github.com/meta-llama/llama3/issues/246}{[llama3\#246]}, \href{https://github.com/QwenLM/Qwen/issues/92}{[Qwen\#92]}). This is followed by \textit{Repetitive Generation} [C.1.2] (46 instances), characterized by repeated phrases or tokens. Other common sub-types include \textit{Special Token Anomaly} [C.1.3] (28 instances), which involves mishandling of tokens like \texttt{[BOS]} or \texttt{[EOS]}, and \textit{Incomplete Response} [C.1.4] (19 instances), which occurs when outputs are prematurely cut off or logically unfinished. Failures to adhere to user-specified constraints are captured as \textit{Not Following Instructions} [C.1.5] (15 instances), including instances like outputting in wrong languages or ignoring formatting requirements (\eg \href{https://github.com/meta-llama/llama/issues/563}{[Llama\#563]}, \href{https://github.com/deepseek-ai/DeepSeek-Coder-V2/issues/12}{[DeepSeek-Coder-V2\#12]}). A key LLM-specific example is \textit{Reasoning Mode Anomaly} [C.1.8] (6 instances), where the LLM fails to correctly enter ``reasoning mode'', leading to incorrect problem-solving approaches (\eg Fig.~\ref{fig:case2}(b)).

A smaller but important category is \textit{Fine-tuning Anomaly} [C.2] (16 instances), which involves unexpected behaviors during model adaptation. \textit{Fine-tuning Convergence Anomaly} [C.2.1] (13 instances) is the primary issue in this category, where the process fails to converge or reaches a suboptimal state (\eg \href{https://github.com/QwenLM/Qwen2.5/issues/935}{[Qwen2.5\#935]}, \href{https://github.com/QwenLM/Qwen/issues/967}{[Qwen\#967]}).


\finding{The prevalence of \textit{Generation Quality Anomaly} (33.9\%) redefines reliability from deterministic crashes to probabilistic cognitive deviations. This shift necessitates moving beyond stack-trace analysis toward semantic behavioral testing to ensure correctness.}\label{fnd:generation_quality}

\subsubsection{Poor Performance [D]}
This category encompasses failures where the LLM operates correctly but unacceptably slowly, impacting its usability.  It is the least common symptom, with 16 reported issues (2.3\%). This category includes \textit{Slow Inference Speed} [D.1] (12 instances), \textit{Slow Fine-tuning Speed} [D.2] (3 instances), and \textit{Slow Loading Speed} [D.3] (1 instance). \textit{Slow Inference Speed} [D.1] is the dominant sub-category, where the model’s response generation time hinders real-time applications (\eg \href{https://github.com/QwenLM/Qwen/issues/105}{[Qwen\#105]}, \href{https://github.com/QwenLM/Qwen/issues/783}{[Qwen\#783]}). 
Fig.~\ref{fig:case2}(c) provides an example, where a user observed a severe performance degradation when running inference inside a Docker container. The model's generation speed was 6 to 7 times slower than when run on the host machine with identical parameters, rendering the deployment unusable for real-time applications.
\textit{Slow Fine-tuning Speed} [D.2] occurs when the fine-tuning process progresses too slowly, while \textit{Slow Loading Speed} [D.3] involves the excessive time required to load the LLM into memory. The low frequency of these issues suggests that performance concerns are less commonly reported as primary symptoms compared to functional or stability failures.

\subsubsection{Comparison with DL System in Symptoms}
\label{subsec:compare_symptom}
We compared our symptom taxonomy with existing studies on traditional DL systems~\cite{DependencyBug2023FSE,fudan2022fse,DLFailure2020ICSE,Junjie2023Tosem,JSFailure2022ASE} to reveal how the LLM paradigm has changed the symptoms of user-encountered failures. The shift from building models from scratch to using large pre-trained generative assets has created entirely new symptoms, amplified the severity of some traditional ones, and diminished others that were once primary concerns.

\textbf{Newly Emerged LLM-Specific Symptoms.} The most profound difference in LLM systems is the emergence of a massive new class of symptoms related to the quality of generated content. \textit{Generation Quality Anomaly} [C.1], which includes symptoms such as \textit{Meaningless Generation}, \textit{Repetitive Generation}, and \textit{Not Following Instructions}, accounts for nearly a third of all failures in our study. This represents a fundamental change in what constitutes a ``bug,'' moving from a verifiable numerical error to a subjective, semantic, and behavioral fault.
Furthermore, the unique architecture and component of LLM systems introduce novel crash symptoms such as \textit{KV Cache Allocation Error} [A.5.3] and \textit{Tokenizer Error} [A.6.2].

\textbf{Amplified Traditional Symptoms.} While \textit{Memory Issue} have always existed~\cite{JSFailure2022ASE,DependencyBug2023FSE}, the revolutionary scale of LLMs has amplified \textit{Memory Issue} into one of the most critical crash types, accounting for 11.2\% of all symptoms. 
The revolutionary increase in model size from megabytes or a few gigabytes to hundreds of gigabytes makes memory management a primary challenge for users.  Similarly, the prevalence and complexity of Tensor-related Errors (12.2\% of symptoms) have been amplified. This is because these failures have shifted from relatively simple dimension mismatches in traditional convolutional layers to the intricate mechanics of the Transformer architecture. Users now frequently encounter more complex and subtle bugs related to attention mask dimensions, sequence length constraints, and numerical instabilities arising from mixed-precision operations.

\textbf{Diminished Traditional Symptoms.} Conversely, the pre-trained paradigm has drastically reduced failures that once plagued DL developers. In studies of DL systems, where users often built components from scratch, \textit{Build Error} was a top-three symptom~\cite{JSFailure2022ASE,Junjie2023Tosem}. In our analysis, this has been reduced to only 4.8\% (a relative decrease of approximately 75\%), reflecting the shift in user behavior away from building and compiling and towards using prepackaged, pre-trained models, effectively abstracting away a major source of user frustration.




\subsection{RQ2: Root Causes of LLM Failures}\label{sec:rq2}
\begin{figure*}[t]
    \centering
    \includegraphics[width=0.9\textwidth]{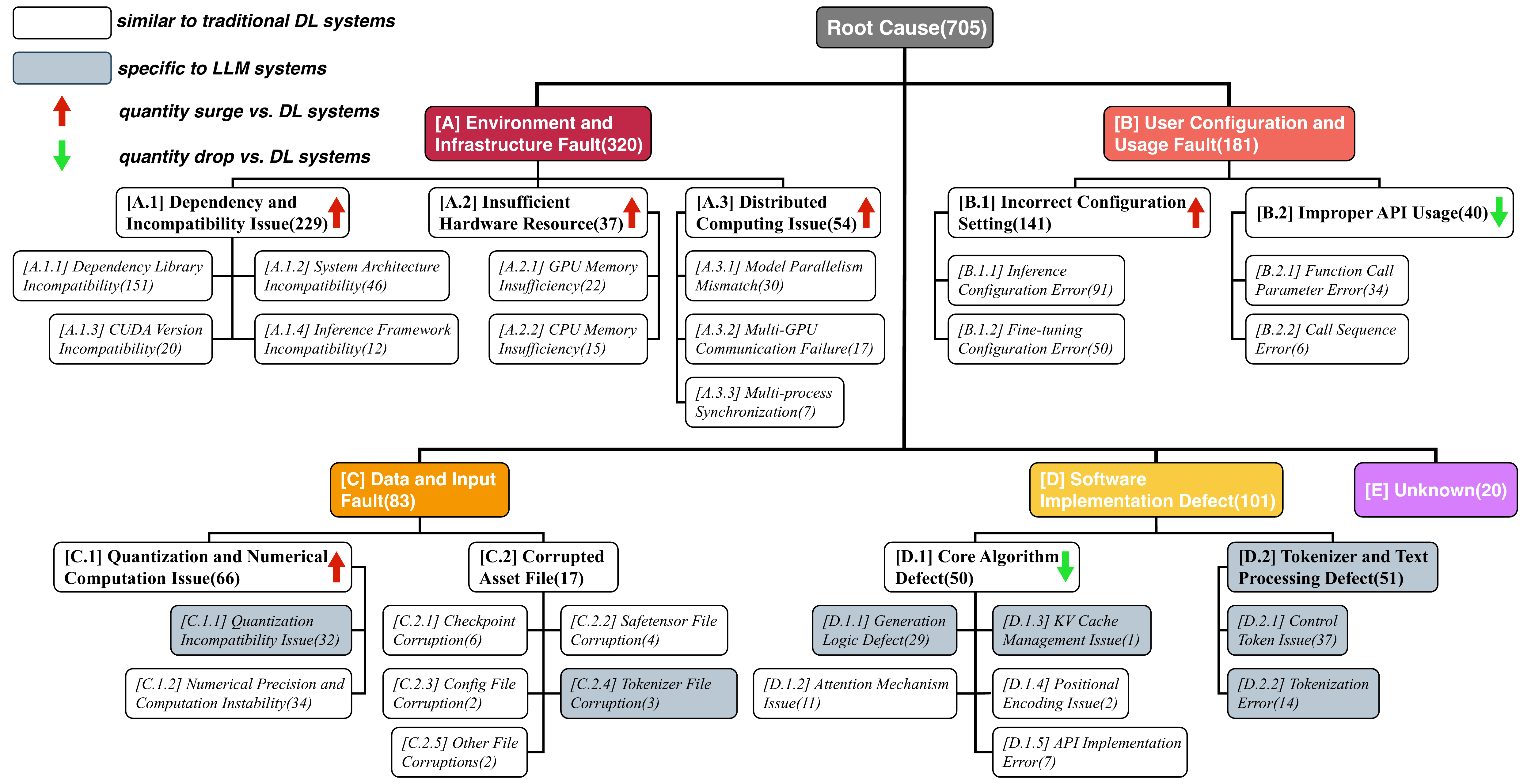}
     \vspace{-0.1in}
    \caption{Root cause taxonomy of failures in open-source LLMs.}
    \label{fig:root_cause_taxonomy}
    \vspace{-0.2in}
\end{figure*}

To address RQ2, we analyze the underlying root causes behind the observed LLM failure symptoms.
Figure~\ref{fig:root_cause_taxonomy} shows the hierarchical taxonomy and distribution of root causes in open-source LLM failures. 
Note that gray rectangles denote LLM-specific root causes (\eg \textit{Control Token Issue}) and white rectangles indicate causes shared with DL systems (\eg \textit{CUDA Version Incompatibility}). 

\subsubsection{Environment and Infrastructure Fault [A]}\label{subsec:rq2_env} 
The most dominant category of root causes is \textit{Environment and Infrastructure Fault} [A], accounting for 320 of 705 instances (45.7\%). This crucial finding reveals that for most users, the primary struggle is not with the LLM's core logic, but with the immense challenge of building a stable and compatible environment in which it can operate.

\textit{Dependency and Incompatibility Issue} [A.1] is the most prevalent group of root causes, with 229 instances. The core problem is \textit{Dependency Library Incompatibility} [A.1.1] (151 instances), where a single incorrect library version can manifest as a wide spectrum of observable symptoms: a \texttt{peft} version mismatch can cause a \textit{Runtime Crash} (Fig.~\ref{fig:case1}(d)), an outdated \texttt{transformers} version can lead to a \textit{Fine-tuning Process Stuck} (Fig.~\ref{fig:case2}(a)), and a faulty \texttt{auto-gptq} library within a Docker image can result in severe \textit{Slow Inference Speed} (Fig.~\ref{fig:case2}(c)). 
Beyond libraries, users frequently struggle with deeper system-level conflicts. \textit{System Architecture Incompatibility} [A.1.2] (46 instances) captures failures arising from mismatches with the host environment, such as unsupported operating systems (\eg Windows or Mac OS) \href{https://github.com/meta-llama/llama/issues/964}{[Llama\#964]}. Similarly, \textit{CUDA Version Incompatibility} [A.1.3] (20 instances) is a critical root cause, occurring when the version of the CUDA toolkit required by the model or its libraries does not match what is installed on the user's system. 
Finally, this root cause also includes \textit{Inference Framework Incompatibility} [A.1.4] (11 instances), such as when an updated model tokenizer becomes incompatible with the vLLM serving framework, leading to malformed output (Fig.~\ref{fig:case2}(b)).

\textit{Insufficient Hardware Resource} [A.2] captures root causes related to physical limitations of user hardware, with 37 instances. The primary cause is \textit{GPU Memory Insufficiency} [A.2.1] (22 instances), which typically manifests as a hard crash when the model's size and KV cache demands exceed the available VRAM \href{https://github.com/QwenLM/Qwen/issues/730}{[QWen\#730]}. Failures from \textit{CPU Memory Insufficiency} (15 instances) are also notable, often manifesting as crashes during the initial model loading phase.
When users attempt to overcome resource limitations by using multiple GPUs, they often encounter a new class of complex root causes, represented by the 54 instances in \textit{Distributed Computing Issue} [A.3]. The most common cause is \textit{Model Parallelism Mismatch} [A.3.1] (30 instances). As an example in Fig.~\ref{fig:case1}(f), a user's script lacking the correct distributed launch parameters (\eg \texttt{--nproc\_per\_node}) will fail when trying to load a multi-GPU checkpoint onto a single-process instance. The other critical causes in this category are \textit{Multi-GPU Communication Failure} [A.3.2] (17 instances) , which points to lower-level networking or NCCL issues, and \textit{Multi-process Synchronization} [A.3.3] (7 instances), which can lead to deadlocks and process hangs \href{https://github.com/meta-llama/llama/issues/704}{[Llama\#704]}.


\finding{The dominance of \textit{Environment and Infrastructure Faults} (45.7\%) reveals the systemic fragility of the LLM software stack. This stems from the rigid coupling between model architectures, hardware drivers, and execution frameworks, where a single version mismatch compromises the entire deployment environment.}\label{fnd:environment_compatibility}

\subsubsection{User Configuration and Usage Fault [B]}\label{subsec:rq2_conf} The second most prevalent category, \textit{User Configuration and Usage Fault} [B] (179 instances, 25.4\%), moves from the environment to the user's direct interaction with the model's interfaces. The high frequency of these faults points to a critical usability challenge in the LLM ecosystem: the configuration and programming interfaces are often overwhelmingly complex and unforgiving, making it easy for even experienced users to make critical mistakes that lead to failure.

\textit{Incorrect Configuration Setting} [B.1] (139 instances) points to failures where users provide invalid parameters for a given fine-tuning or inference task.
\textit{Inference Configuration Error} [B.1.1] is the most frequent sub-type, with 91 instances. These failures occur when users misconfigure crucial parameters related to the generation process or resource management. A typical example in Fig.~\ref{fig:case1}(b) is when a user fails to set a correct value for \texttt{max\_model\_len}, leading to a crash when the model's sequence length exceeds the allocated KV Cache size. 
\textit{Fine-tuning Configuration Error} [B.1.2] (50 instances) present another major hurdle during fine-tuning. These faults are typically caused by the immense complexity of correctly configuring distributed training parameters for frameworks such as DeepSpeed~\cite{DeepSpeed} or FSDP~\cite{FSDP} (\eg \href{https://github.com/QwenLM/Qwen3/issues/306}{[QWen3\#306]}). Users often struggle to navigate the intricate setup for multi-GPU fine-tuning, leading to errors that halt the model adaptation process.

\textit{Improper API Usage} [B.2] captures more direct coding mistakes when interacting with the model's API, accounting for 40 instances. The vast majority are \textit{Function Call Parameter Error} [B.2.1] (34 instances). These are structural errors in how functions are called. The example in Fig.~\ref{fig:case1}(e) where the wrong number of positional arguments is provided to a model's \texttt{\_\_init\_\_function}, is a clear illustration of this issue. A smaller but distinct group is \textit{Call Sequence Error} [B.2.2] (6 instances). This error occurs when API functions are called in an illogical order that violates the API's expected workflow. Ultimately, the prevalence of these faults underscores that an LLM system's usability is defined as much by the clarity of its configuration and API design as it is by its generative power.

\finding{The prevalence of \textit{User Configuration Faults} (25.4\%) exposes a critical usability deficit. Current frameworks expose complex hyperparameter spaces without sufficient defensive design or runtime validation, allowing minor configuration errors to propagate into fatal system failures.}\label{fnd:configuration}

\subsubsection{Data and Input Fault [C]}\label{subsec:rq2_data} This category, with 83 instances (11.8\%), encompasses failures originating not from the code or environment, but from the data assets the model consumes. The majority of these issues fall under \textit{Quantization and Numerical Computation Issue} [C.1] (66 instances), a category that is a hallmark of the modern LLM era. To run massive models on consumer hardware, users are practically required to use low-precision data formats, forcing them to navigate a volatile and immature ecosystem. 
\textit{Quantization Incompatibility Issue} [C.1.1] is a major hurdle for users, accounting for 32 instances. As shown in Fig.~\ref{fig:case1}(c), these failures typically arise when a model's configuration specifies a quantization method (\eg \texttt{fp8}) that is not supported by the user's execution framework, which may only support types like \texttt{awq}~\cite{MLSYS2024AWQ} or \texttt{gptq}~\cite{GPTQ}. \textit{Numerical Precision and Computation Instability} [C.1.2] (34 instances) represents a more subtle class of data-related bugs. These failures can arise when using low-precision data types or specific inference parameters that lead to numerical errors like \texttt{NaN} or \texttt{inf}. As shown in the example in Fig.~\ref{fig:case1}(a), using a very low \texttt{temperature} setting with a quantized model can cause a softmax operation to overflow, resulting in a \texttt{RuntimeError}.
\textit{Corrupted Asset File} [C.2], with 17 instances, represents more traditional data integrity problems where a critical file is damaged or malformed, typically leading to a crash during the model loading phase. The failures are distributed across a range of essential files, including model weights (\textit{Checkpoint Corruption}, \textit{Safetensor File Corruption}), model configurations (\textit{Config File Corruption}) and tokenizer file (\textit{Tokenizer File Corruption}).


\finding{The emergence of \textit{Quantization Issues} (9.3\%) highlights a critical trade-off: aggressive quantization enables hardware accessibility but introduces numerical instability. This challenge is exacerbated by the lack of standardization in quantization formats, shifting the burden of precision management from developers to end-users.}\label{fnd:quantization}

\subsubsection{Software Implementation Defect [D]}\label{subsec:rq2_software} This category, with 101 instances (14.3\%), captures failures caused by traditional software bugs in the source code, its tokenizer or the surrounding frameworks. \textit{Core Algorithm Defect} [D.1] accounts for 50 instances and represents bugs in the logic of the LLM itself.  The most common type in [D.1] is \textit{Generation Logic Defect} [D.1.1] (29 instances), which are bugs in the code that controls the text generation process, such as the sampling methods or repetition penalties. These defects are a direct cause of observable symptoms such as repetitive or meaningless output. Other defects are found in the model's fundamental implementation, including the management of the KV Cache (1 instance) and Positional Encoding (2 instances). Beyond logic defect, \textit{Tokenizer and Text Processing Defect} [D.2] (51 instances), captures bugs in the crucial front-end component that processes natural language input and output.
The vast majority of these are \textit{Control Token Issue} [D.2.1] (37 instances), where the software incorrectly handles the special tokens that structure a conversation (\eg \texttt{[BOS]}, \texttt{[EOS]}, or tokens used in chat templates). A bug here can lead to malformed output or cause the model to misunderstand the conversational context. The remaining \textit{Tokenization Error} [D.2.2] (14 instances) are defects in the tokenization algorithm itself, corrupting the model's input from the very start.

\subsubsection{Unknown [E]} \textit{Unknown} [E] category (20 instances, 2.8\%) includes cases where a definitive root cause could not be determined despite a meticulous review of the GitHub issue. This typically occurred when a user's report and the subsequent discussion lacked sufficient diagnostic information to pinpoint the failure's origin. 

\subsubsection{Comparison with DL System in Root Cause}
We compared our root cause taxonomy with existing studies on traditional DL systems~\cite{DependencyBug2023FSE,fudan2022fse,DLFailure2020ICSE,Junjie2023Tosem,JSFailure2022ASE} to reveal how the LLM paradigm has changed the root causes of user-encountered failures. This comparison reveals that the LLM paradigm has triggered a fundamental paradigm shift in the challenges facing developers, moving them away from the complexities of model creation and toward the frictions of system integration. 

\textbf{Newly Emerged LLM-Specific Root Causes.}
The modern LLM ecosystem has introduced specialized components and practices that have become dominant sources of failure. 
While the concept of model quantization is not new, the necessity of running massive models on consumer hardware has transformed it from a niche optimization into a mainstream, error-prone process. This has created a volatile and fragmented ecosystem of competing, often incompatible formats (\eg \texttt{awq} or \texttt{gptq}), making \textit{Quantization Incompatibility} (32 instances) a novel and critical source of user failure.
Furthermore, our study identifies a major new category of 51 failures originating in the tokenizer and its surrounding text processing logic. These include bugs in handling \textit{Control Token Issue} (37 instances) and \textit{Tokenization Error} (14 instances). These are entirely novel root causes, as complex tokenization and chat templates have no parallel in DL systems.

\textbf{Amplified Traditional Root Causes.} Some classic DL system problems have been dramatically amplified by the scale and complexity of the LLM stack, transforming them from secondary concerns into the most dominant root causes of failure. The clearest example is \textit{Dependency and Incompatibility Issue}, which at 227 instances, represents an explosive growth compared to DL studies~\cite{JSFailure2022ASE,Junjie2023Tosem}. The modern LLM software stack is a fragile house of cards, requiring users to perfectly coordinate specific versions of numerous libraries (transformer, PEFT, quantization engines, and inference frameworks). A single version mismatch can cause the entire system to collapse.


\textbf{Diminished Coding-Centric Failures.} Conversely, the shift to pre-trained assets has drastically reduced failures related to Model Construction. In traditional DL studies~\cite{DLFailure2020ICSE}, a major root cause was dimension mismatch errors inside manually written neural network layers (\eg inside forward() functions). In the LLM paradigm, these faults have diminished, as users now primarily orchestrate pre-packaged black-box models rather than implementing neural architectures from scratch.



\begin{table*}[t]
\caption{Failure distribution by symptom for each sub-root-cause}
\vspace{-0.1in}
\label{tab:relation} 
\centering
\renewcommand{\arraystretch}{0.9}
\resizebox{0.95\textwidth}{!}{%
\begin{tabular}{@{}l|l||c|c|c|c|c@{}}
\toprule
\multicolumn{2}{c||}{\diagbox{\textbf{Root Cause}}{\textbf{Symptom}}} & \makecell{\textbf{Runtime} \\ \textbf{Crash}} & \makecell{\textbf{Process} \\ \textbf{Hang}} & \makecell{\textbf{Incorrect} \\ \textbf{Functionality}} & \makecell{\textbf{Poor} \\ \textbf{Performance}} & \textbf{Total} \\
\midrule
\multirow{3}{*}{\textbf{Environment and Infrastructure Fault}} & Dependency and Incompatibility Issue & 149 & 4 & 67 & 9 & \multirow{3}{*}{\textbf{320}} \\
& Insufficient Hardware Resource & 36 & 0 & 1 & 0 & \\
& Distributed Computing Issue & 42 & 8 & 1 & 3 & \\
\midrule
\multirow{2}{*}{\textbf{User Configuration and Usage Fault}} & Incorrect Configuration Setting & 72 & 4 & 63 & 2 & \multirow{2}{*}{\textbf{181}} \\
& Improper API Usage & 32 & 0 & 7 & 1 & \\
\midrule
\multirow{2}{*}{\textbf{Data and Input Fault}} & Corrupted Asset File & 15 & 0 & 2 & 0 & \multirow{2}{*}{\textbf{83}} \\
& Quantization and Numerical Computation Issue & 31 & 0 & 35 & 0 & \\
\midrule
\multirow{2}{*}{\textbf{Software Implementation Defect}} & Core Algorithm Defect & 25 & 2 & 22 & 1 & \multirow{2}{*}{\textbf{101}} \\
& Tokenizer and Text Processing Defect & 10 & 0 & 41 & 0 & \\
\midrule
\multirow{1}{*}{\textbf{Unknown}} & - & 19 & 1 & 0 & 0 & \multirow{1}{*}{\textbf{20}} \\
\midrule
\multicolumn{2}{l||}{\textbf{Total}} & \textbf{431} & \textbf{19} & \textbf{239} & \textbf{16} & \textbf{\num} \\
\bottomrule
\end{tabular}}
\vspace{-0.1in}
\end{table*}

\subsection{RQ3: Relation between Symptoms and Root Causes}\label{sec:rq3} 
By mapping the observable symptoms from RQ1 (\S~\ref{sec:rq1}) to the underlying root causes from RQ2 (\S~\ref{sec:rq2}), we can construct a diagnostic map for troubleshooting LLM failures. This user-centric analysis, summarized in Table~\ref{tab:relation}, reveals distinct data-driven heuristics to pinpoint the likely origin of a problem based on what the user observes.

\textbf{When the Program Crashes.} Our data shows that a \textit{Runtime Crash} (431 instances) is overwhelmingly rooted in the user's environment or configuration, rather than a bug in the model's core logic. 
\textit{Environment and Infrastructure Fault} is the primary source of crashes, accounting for over half of all instances (52.9\%) . More specifically, \textit{Dependency and Incompatibility Issue} is the single most probable cause of any given crash (34.6\% of the time). This provides a clear, actionable heuristic for users: when faced with a crash, the first places to investigate are the library dependencies and configuration files.

\textbf{When the Output is Incorrect.} For \textit{Incorrect Functionality} failures (239 instances), environmental and configuration issues remain the most probable root causes, just as with crashes. However, this scenario introduces a crucial new diagnostic possibility: \textit{Tokenizer and Text Processing Defect} becomes a major and highly specific potential cause, accounting for a substantial 17.2\% of these semantic failures. While these internal software bugs rarely cause a full system crash, they are a leading source of malformed or non-sensical output, offering a distinct path for investigation when the system is otherwise stable.

\textbf{When the Program Hangs or is Slow.}
For less common but critical operational failures, the root causes are highly specific. \textit{Process Hang} (19 instances) is strongly linked to the complexities of multi-GPU setups, with \textit{Distributed Computing Issue} being the cause 42.1\% of the time (8 of 19 instances). \textit{Poor Performance} (16 instances), on the other hand, is most often a manifestation of a broken software stack; \textit{Dependency and Incompatibility Issue} is the root cause in 50\% of all slowdowns, likely due to misconfigured libraries preventing hardware acceleration.


\finding{Symptom-cause mapping reveals a diagnostic divergence: \textit{Runtime Crash} serves as predominant indicators of environmental and configuration faults, whereas \textit{Incorrect Functionality} acts as a distinctive signature for tokenizer defects. This suggests prioritized workflows: environment stabilization for crashes versus data pipeline auditing for semantic anomalies.}\label{fnd:relation}


\subsection{RQ4: A Comparative Analysis across LLM Series}\label{sec:rq4}

\begin{figure}[t]
	\centering
	\subfigure[Symptom]{
		\begin{minipage}[b]{0.48\linewidth}
			\includegraphics[width=1\textwidth]{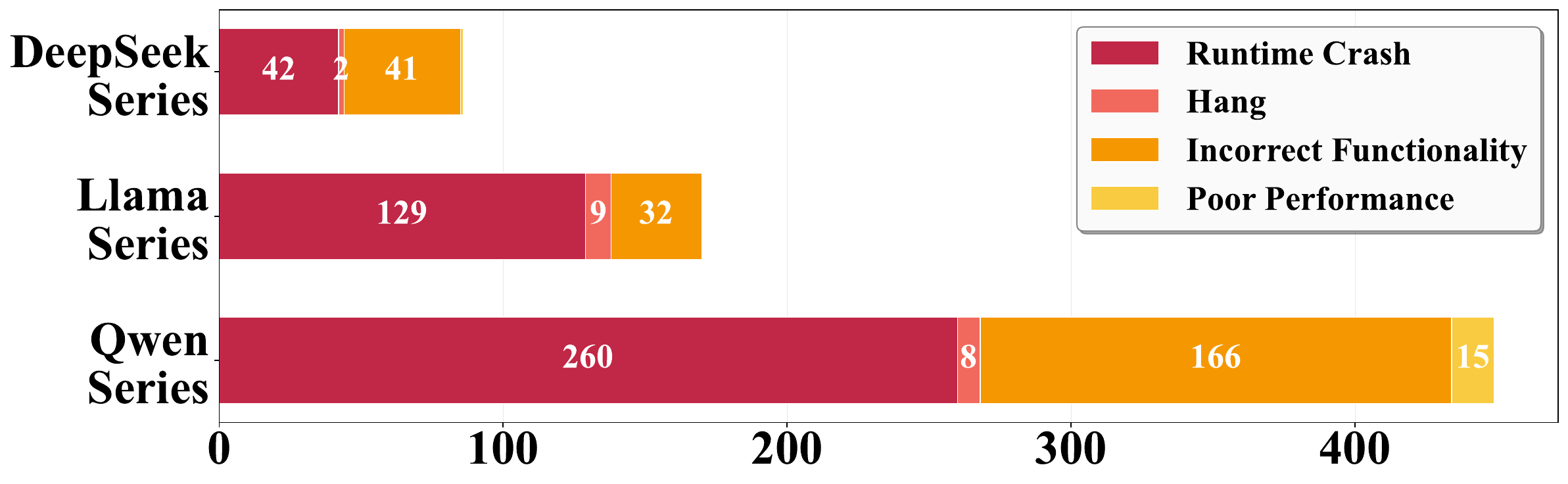}
		\end{minipage}
		\label{fig:symptom_distribution}
	}
    \subfigure[Root Cause]{
    	\begin{minipage}[b]{0.48\linewidth}
   		\includegraphics[width=1\textwidth]{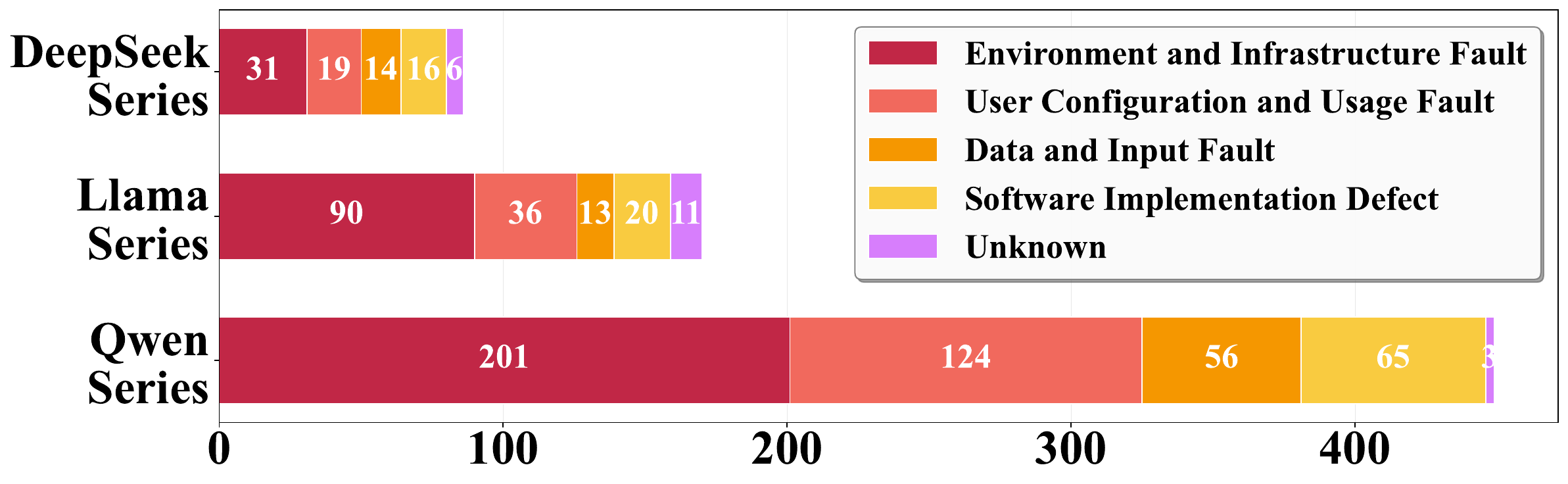}
    	\end{minipage}
	\label{fig:rootcause_distribution}
    }
        \vspace{-0.2in}
	\caption{Failure distribution across DeepSeek, Llama, Qwen series.}
	\label{fig:distribution}
    \vspace{-0.3in}
\end{figure}

A crucial question for our study is whether the identified failure patterns are model-specific quirks or if they generalize across the open-source landscape. To address RQ4, we conducted a comparative analysis of failure patterns across the DeepSeek, Llama, and Qwen series (Fig.~\ref{fig:distribution}), investigating the generalizability of both what users see (symptoms) and why failures occur (root causes).

\textbf{Symptoms: A Partially Generalizable User Experience.}
At the symptom level, the user experience is only partially generalizable. As Fig.~\ref{fig:symptom_distribution} shows, the dominant symptoms (\ie \textit{Runtime Crash} and \textit{Incorrect Functionality}) are consistent across all three series, indicating a broadly shared user struggle. However, the specific distribution of these symptoms does not fully generalize. Llama users report four times as many crashes as functional issues, while for DeepSeek the split is nearly even. This divergence suggests that a symptom-level view alone is insufficient to understand the core problem and find the truly generalizable pattern. Consequently, we need to analyze the underlying root causes.

\textbf{Root Causes: High Generalizability Reveals a Systemic Problem.}
In stark contrast to the symptoms, the root causes of failure exhibit a high degree of generalizability across all three series. Figure~\ref{fig:rootcause_distribution} reveals that the overwhelming majority of failures in Llama (74\%), Qwen (73\%), and DeepSeek (58\%) stem from the identical two categories: \textit{Environment and Infrastructure Fault} and \textit{User Configuration and Usage Fault}. The strong generalizability of these external root causes is the most compelling evidence of our comparative analysis. It provides powerful evidence that the primary reliability challenges are not idiosyncratic to the model series, but are systemic to the complex deployment and configuration ecosystem that all these model series share.



\finding{A cross-series comparison reveals a duality: while failure symptoms vary locally, underlying root causes exhibit systemic homogeneity. The universal dominance of infrastructure and configuration faults across all series confirms that reliability challenges are inherent to the shared ecosystem, rather than being idiosyncratic quirks of specific model architectures.}\label{fnd:comm}

\begin{figure}[t]
	\centering
	\subfigure[Symptom]{
		\begin{minipage}[b]{0.48\linewidth}
			\includegraphics[width=1\textwidth]{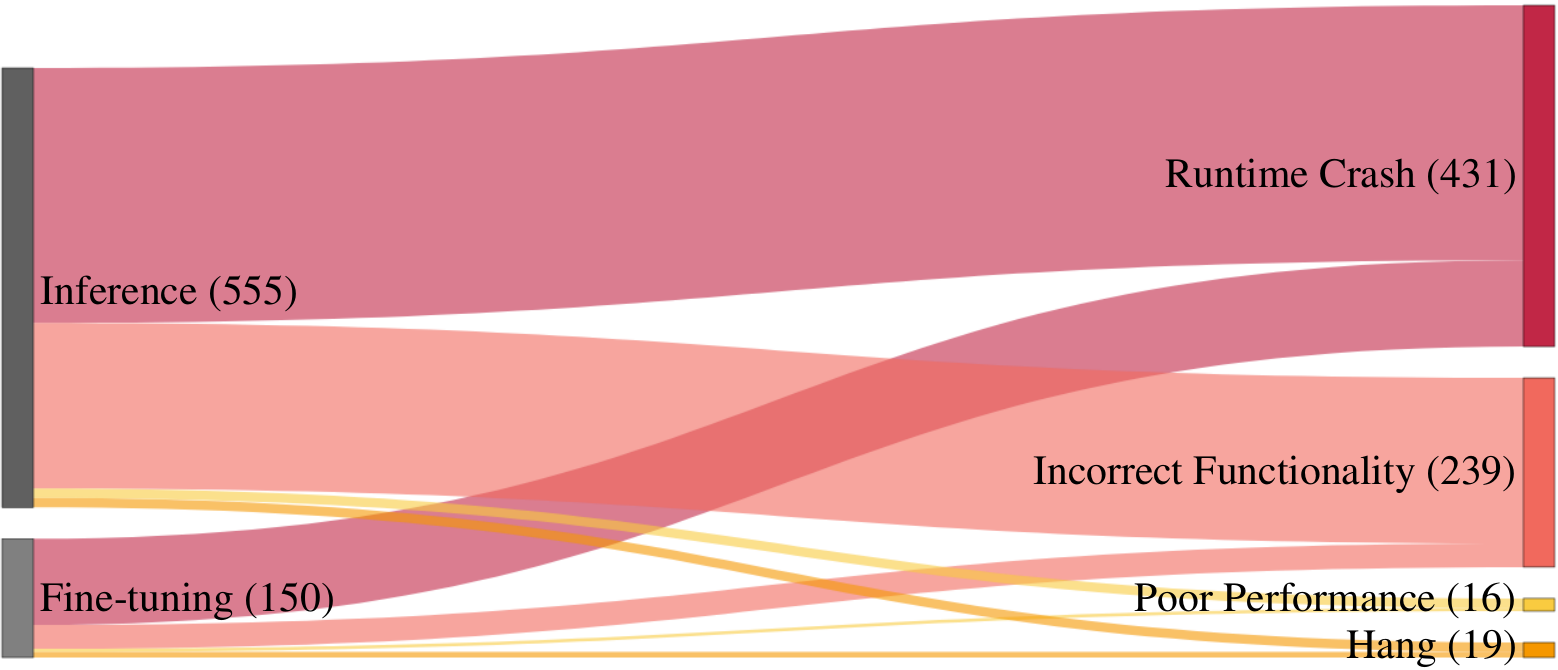}
		\end{minipage}
		\label{fig:symptom_phase}
	}
    \subfigure[Root Cause]{
    	\begin{minipage}[b]{0.48\linewidth}
   		\includegraphics[width=1\textwidth]{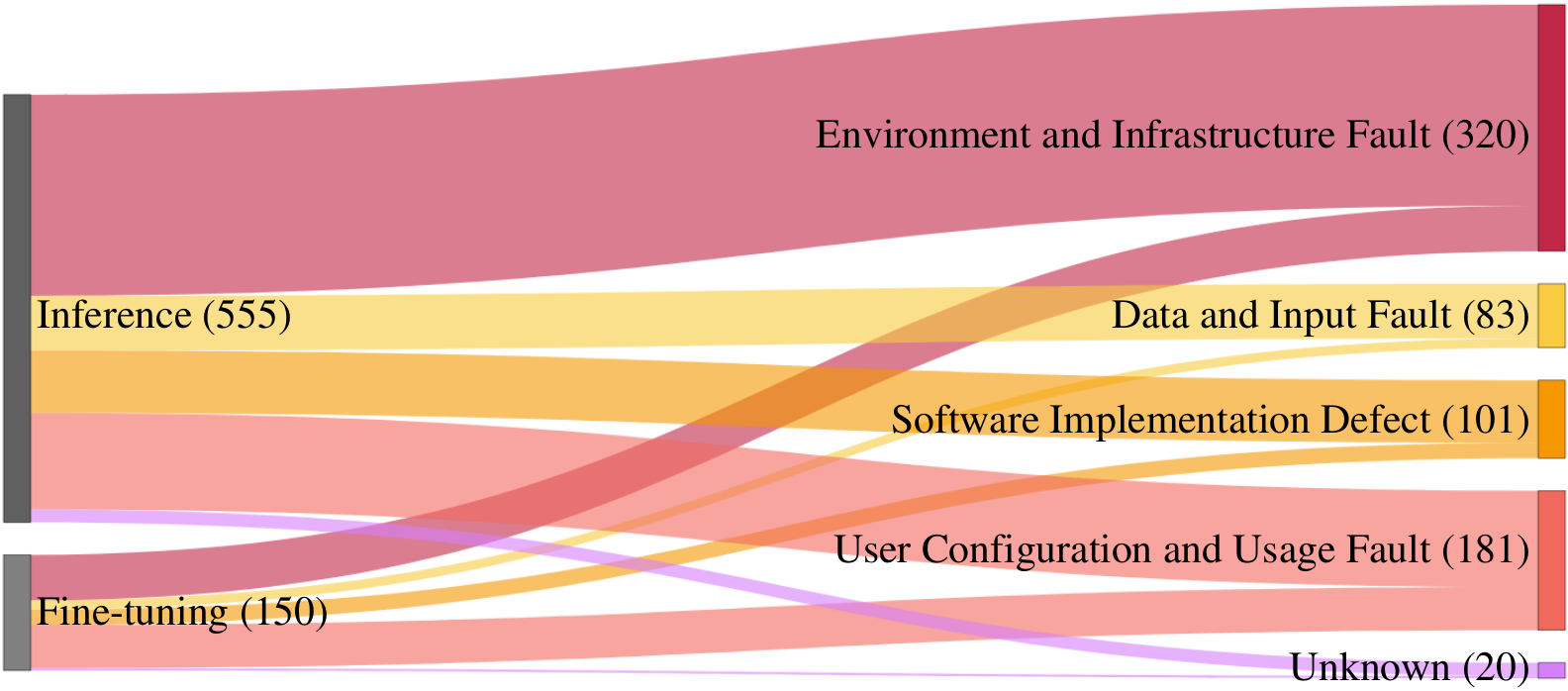}
    	\end{minipage}
	\label{fig:rootcause_phase}
    }
    \vspace{-0.2in}
	\caption{The lifecycle-dependent shift in failure patterns (Flow thickness corresponds to issue volume).}
	\label{fig:rq5-phase}
    \vspace{-0.2in}
\end{figure}

\subsection{RQ5: Stage-Specific Failure Patterns}\label{sec:rq5}

To understand how user challenges evolve throughout the LLM lifecycle, we analyzed the transition of failure patterns between fine-tuning (150 issues) and inference (555 issues) (Fig.~\ref{fig:rq5-phase}). This analysis reveals a dramatic structural shift, indicating that users face fundamentally different reliability barriers at each stage.

Figure~\ref{fig:symptom_phase} shows that the symptoms experienced during fine-tuning are overwhelmingly dominated by Runtime Crashes. The thick flow from ``Fine-tuning'' to ``Runtime Crash'' illustrates that users in this phase are primarily engaged in a battle for basic system stability (\ie simply getting the complex training process to run without collapsing).
In the inference stage, the landscape of symptoms changes significantly. While \textit{Runtime Crash} remain the most frequent single symptom, \textit{Incorrect Functionality} surges to become a nearly equal concern. This prominent flow from ``Inference'' to ``Incorrect Functionality'' highlights a crucial shift in the user's focus: once a model is stable enough to run, the primary challenge becomes managing the quality of its behavioral output.

Figure~\ref{fig:rootcause_phase} reveals the underlying reasons for this change in symptoms. Failures during fine-tuning are predominantly caused by \textit{User Configuration and Usage Fault}. This strong flow reflects the immense complexity of correctly configuring distributed training parameters and the intricate software stack required for model adaptation, making user error the primary driver of failures.
Conversely, failures during inference are dominated by \textit{Environment and Infrastructure Fault}. This decisive shift underscores the challenges of deploying a model into diverse, real-world production or local environments. At this stage, the primary drivers of failure are no longer the user's initial setup choices, but rather the friction and incompatibilities that arise from the surrounding hardware, dependency libraries, and the broader software stack.


\finding{The transition from fine-tuning to inference represents an escalation rather than a substitution of failure modes. While \textit{User Configuration Faults} remain a persistent hurdle, the Inference stage compounds this challenge with a massive influx of \textit{Environment and Infrastructure Faults}, forcing users to battle ecosystem incompatibilities alongside configuration struggles.}\label{fnd:stage}

\subsection{RQ6: Evolution of LLM Failures}\label{sec:rq6}

\begin{figure}[t]
	\centering
	\subfigure[Symptom]{
		\begin{minipage}[b]{0.48\linewidth}
			\includegraphics[width=1\textwidth]{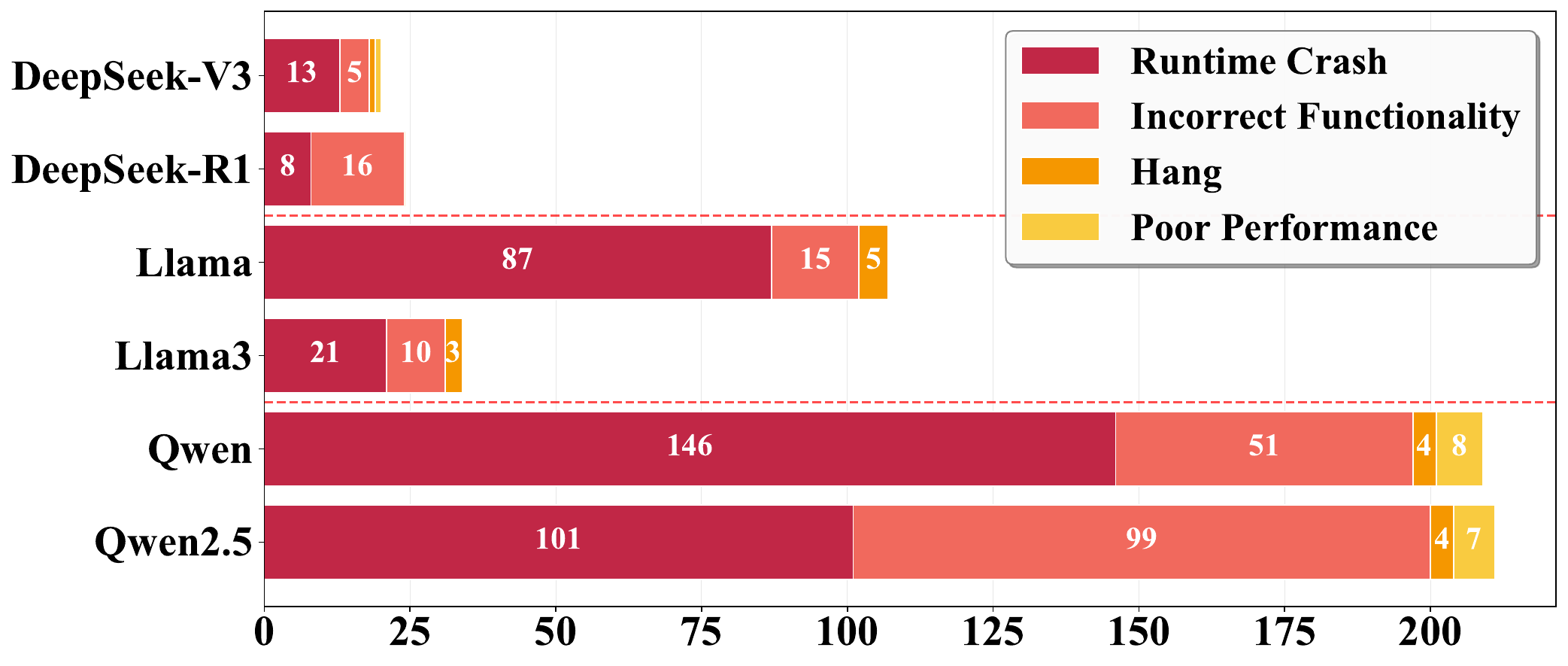}
		\end{minipage}
             
		\label{fig:symptom_evolution}
	}
     \vspace{-0.2in}
    \subfigure[Root Cause]{
    	\begin{minipage}[b]{0.48\linewidth}
   		\includegraphics[width=1\textwidth]{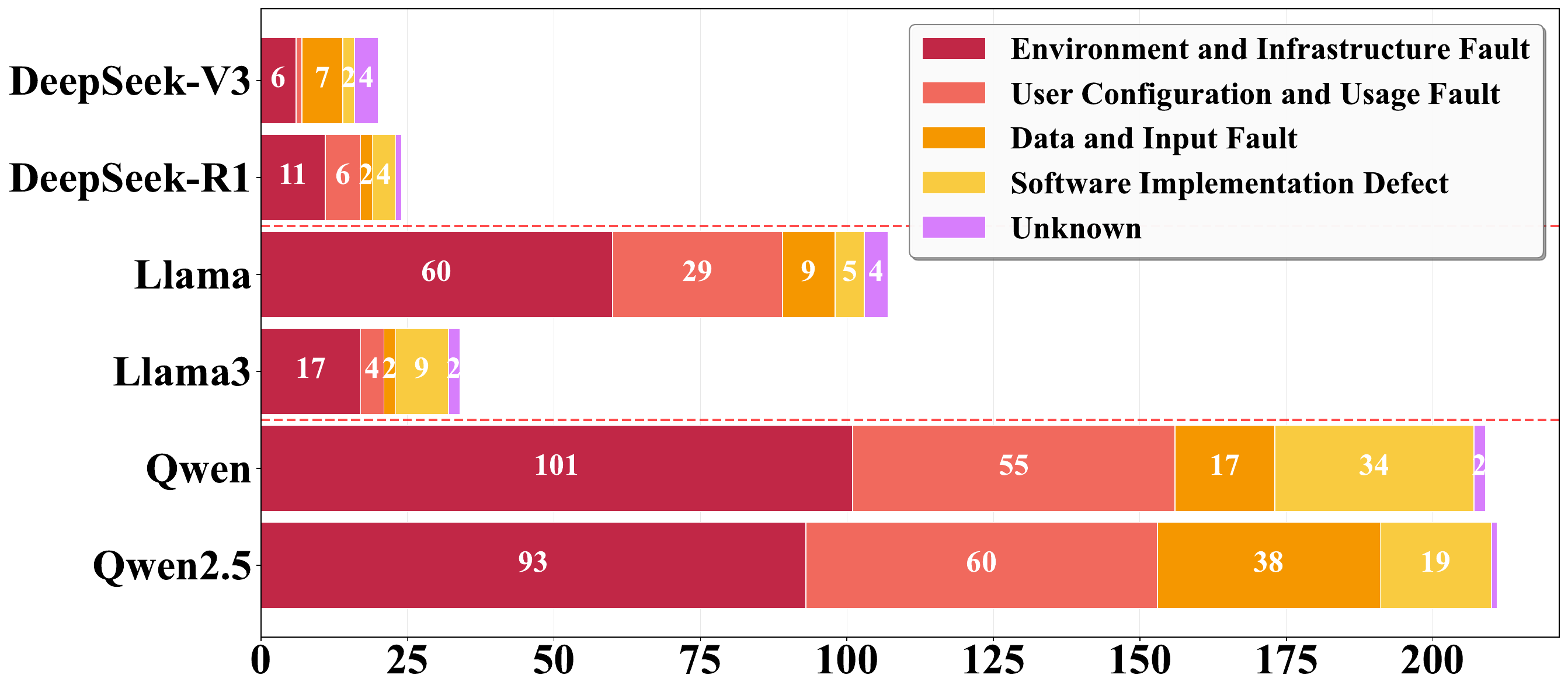}
    	\end{minipage}
	\label{fig:rootcause_evolution}
    }
	\caption{Failure distribution across LLMs in the same series.}
	\label{fig:evolution}
    \vspace{-0.2in}
\end{figure}

To understand the reliability trajectory of the rapidly evolving LLM ecosystem, we analyzed how failure patterns change between successive model versions within the same series (RQ6). Our findings, illustrated in Figure~\ref{fig:evolution}, reveal a clear maturation pattern: as models and their surrounding tooling improve, the nature of user challenges evolves from foundational stability to advanced implementation and optimization issues.

\textbf{Symptom Evolution.} We observe a consistent and significant shift in dominant symptoms, shifting from Operational Viability (making it run) to Semantic Fidelity (making it behave). In the Qwen series, the older Qwen was dominated by \textit{Runtime Crash} (146 of 245 issues, or 59.6\%). In the newer Qwen2.5, while crashes are still common, \textit{Incorrect Functionality} surged to become almost equally prevalent (99 of 247 issues, or 40.1\%, up from 20.8\% in the original). The Llama series shows a similar trend, along with a notable 68\% reduction in total reported issues from Llama 2 to Llama 3, signaling significant maturation. The share of \textit{Runtime Crash} decreased (from 81.3\% to 61.8\%), while the proportion of \textit{Incorrect Functionality} more than doubled (from 14.0\% to 29.4\%). 
This consistent trend indicates that as providers solve fundamental stability problems, the focus of user-reported failures naturally shifts to the more complex and subtle challenge of ensuring the model's behavioral correctness and output quality.



\textbf{Root Cause Evolution.} The underlying root causes also show a distinct evolutionary pattern, which can be seen as three case studies in ecosystem maturation. In the Qwen series, the most notable change is the doubling of \textit{Data and Input Fault} (from 8.1\% to 18.1\%). This is likely driven by the increased complexity and user adoption of quantization in Qwen2.5, highlighting how new optimization techniques introduce new classes of failure. The Llama series demonstrates a dramatic improvement in user experience. \textit{User Configuration and Usage Fault} is more than halved (from 27.1\% to 11.8\%), suggesting that Llama 3 is significantly easier to configure. Interestingly, this exposed deeper issues, as the proportional share of \textit{Software Implementation Defect} increased (from 4.7\% to 26.5\%). As the ``easy'' configuration problems were solved, the remaining users were more likely to encounter and report the ``hard'' fundamental bugs in the core software. The DeepSeek series shows that progress can come with a cost. As models become more powerful (DeepSeek-R1 vs. V3), they place greater demands on the user's system, causing the proportion of \textit{Environment and Infrastructure Fault} to grow significantly (from 31.6\% to 45.8\%).


\finding{As the ecosystem matures, the failure landscape evolves distinctively: symptoms shift from foundational stability issues (Crashes) to subtle semantic anomalies, while root causes migrate from basic usability friction towards the structural complexities of advanced optimization and environmental scale.}\label{fnd:evolution}




\section{Discussion}

\subsection{Actionable Lessons from Failure Patterns}
\label{sec:lesson}
Our empirical analysis offers a series of lessons for model providers and practitioners. We structure these lessons around the key findings of our study, addressing the novel failures unique to LLMs, the traditional failures amplified by scale, and the strategic insights from broader failure patterns.

\textbf{Lessons from Novel LLM-Specific Failures}
The emergence of LLMs has introduced entirely new classes of failures that demand a shift in software engineering practices. Our finding that \textit{Generation Quality Anomaly} is a dominant symptom (Finding~\ref{fnd:generation_quality}), compounded by failures in non-code assets like tokenizers (\S~\ref{subsec:rq2_data}), reveals that the definition of a ``bug'' has expanded to include non-deterministic semantic faults. 
This requires model providers to move beyond traditional testing to develop extensive behavioral test suites. For instance, Llama 3 repository now includes a dedicated \texttt{test\_tokenizer.py} suite to validate special token encodings independently of model weights~\cite{Llama3Test}. Qwen2.5-Coder employs executor-based validation to filter synthetic training data~\cite{qwen25codertechnicalreport}. 

In addition, the emergence of \textit{Quantization and Numerical Computation Issue} as a major failure category (Finding~\ref{fnd:quantization}) highlights a critical trade-off between hardware accessibility and reliability. This reveals an urgent need for model providers to address the immaturity of the deployment ecosystem by releasing official pre-quantized model variants. DeepSeek-V3 addresses this by implementing specialized FP8 kernels with blockwise scaling to prevent numerical overflow~\cite{deepseekv3technicalreport}, and Qwen2.5 releases official AWQ/GPTQ variants validated against internal calibration datasets~\cite{QwenCollections}. Until the ecosystem matures, practitioners should treat quantization as a high-risk procedure, prioritizing officially validated models over ad-hoc conversions.

\textbf{Lessons from Amplified Traditional Failures} 
The immense scale of LLMs has amplified traditional software engineering challenges into primary barriers for users. The dominance of \textit{Environment and Infrastructure Fault} (Finding~\ref{fnd:environment_compatibility}) proves that the deployment stack is a greater obstacle for users than the model itself. This teaches a clear lesson for model providers: the user's deployment experience is as critical as model performance. Providers can dramatically reduce these failures by transitioning to standardized distribution stacks, such as Llama Stack~\cite{Llamastack}, which provides prepackaged Docker distributions to encapsulate complex dependencies.

Beyond the software stack, the models' scale creates intense pressure on user hardware, leading to resource and distributed computing failures. For practitioners, the lesson is that computational efficiency is a core reliability practice. They should adopt Parameter-Efficient Fine-Tuning~\cite{PEFT} to minimize the hardware footprint and use high-level orchestration frameworks (\eg vLLM~\cite{vllm}, DeepSpeed~\cite{DeepSpeed}) to abstract away the error-prone complexities of multi-GPU communication.

\textbf{Lessons from Broader Failure Patterns} 
Our analysis of cross-cutting patterns teaches a crucial strategic lesson: LLM reliability is not a static property but a dynamic one, highly dependent on both the user's task and the system's maturity. This context-dependency is evident across the different lifecycle stages (Finding~\ref{fnd:stage}). The dominance of \textit{User Configuration and Usage Faults} during fine-tuning highlights a preventable pain point (Finding~\ref{fnd:configuration}). This presents a clear opportunity for framework developers to provide stage-aware support, such as ``dry-run'' features. A prime example is PyTorch's torchtitan, which implements a \texttt{fake\_backend} to validate distributed configurations without allocating costly GPU resources~\cite{Dryrun}.

Reliability is also dynamic over time, as our finding on failure pattern evolution proves (Finding~\ref{fnd:evolution}). As models mature, so do their failure modes, meaning reliability practices cannot be static. This suggests a best practice for providers: treat reliability as a living concern by adopting longitudinal tracking. The release of DeepSeek-V3.1-Terminus~\cite{DeepSeek-V3.1-Terminus}, which specifically targeted community-reported stability issues and documented known precision limitations, serves as a model for helping users navigate a constantly moving target.

\subsection{Threats to Validity}
\label{sec:threats}

\textbf{External Validity.} A potential threat to external validity is whether our findings, derived from open-source LLMs, can be generalized to closed-source models (\eg OpenAI's GPT-4). We argue that our findings are highly relevant to these systems for two primary reasons: (1) Shared Foundational Architectures: The leading closed-source models are built on the same foundational architectures (\eg Transformer~\cite{transfomer} and Mixture-of-Experts (MoE)~\cite{MoE})  as the open-source models in our study. As many failures are intrinsically tied to these core architectural principles (\eg \textit{Attention Mechanism Issues} and \textit{Quantization Incompatibility Issues}), our findings in these areas are likely transferable. (2) Convergent Technology Stacks: In production, both open- and closed-source models rely on a convergent technology stack, including GPUs with CUDA, frameworks like PyTorch, and high-throughput inference engines like vLLM. Consequently, our findings on LLM failures (\eg \textit{CUDA Version Incompatibility} and \textit{Checkpoint Corruption}) are directly relevant to the operational challenges faced by closed-source providers.
While our findings are most applicable to the open-source ecosystem, the shared architectural and infrastructural foundations suggest that they provide a strong indicative model for reliability challenges in closed-source systems. We acknowledge that our dataset does not include failures from proprietary models, which remains an avenue for future work.

\textbf{Internal Validity.} The manual labeling of \num GitHub issues to categorize failure symptoms, root causes, and lifecycle stages introduces potential subjectivity. To mitigate this, we employed a rigorous multi-stage protocol grounded in established qualitative research practices (Section~\ref{sec:data_labeling}). The key quality control measures included:
(1) Independent dual-labeling of every issue by two experienced researchers. (2) A systematic process for resolving discrepancies through consensus discussion, with a senior author serving as a final arbiter for complex cases. (3) A team-wide calibration process using a detailed codebook to ensure a shared and consistent understanding of the taxonomy.
The reliability of this process was quantitatively validated by a high inter-annotator agreement score (Cohen's Kappa $> 0.90$), confirming a high degree of consistency and minimizing the impact of subjective bias in our categorizations.

\subsection{Validation with Industrial Reliability Initiatives}
We cross-referenced our results with recent engineering releases and technical reports from major open-source model providers. The industry's independent engineering efforts strongly corroborate our failure categorization.

\textbf{Validation of Infrastructure Friction (Finding~\ref{fnd:environment_compatibility}).}
We identified \textit{Environment and Infrastructure Faults} as the leading cause of crashes (45.7\%), highlighting the brittleness of the deployment stack. This bottleneck has directly driven the industry toward containerized standardization. 
For instance, Meta recently introduced the Llama Stack~\cite{Llamastack} to provide prepackaged verified distributions via Docker, specifically designed to encapsulate complex dependencies and prevent the ecosystem-driven failures we observed. Similarly, hardware vendors have validated this need by releasing specialized Docker images for non-standard hardware, such as AMD's official ROCm containers for vLLM~\cite{ROCm}  and Intel's IPEX-LLM containers~\cite{ipex-llm}. These initiatives confirm that resolving environment friction is currently a top-priority engineering goal in the ecosystem.

\textbf{Validation of Non-Deterministic Output Management (Finding~\ref{fnd:generation_quality}).}
Our study highlights \textit{Generation Quality Anomaly} as a dominant symptom, necessitating new forms of validation. This is validated by the widespread deployment of external safety and structural guardrails. 
The release of Llama Guard 3~\cite{Llama-Guard-3-8B} and QwenGuard~\cite{zhao2025qwen3guardtechnicalreport}  confirms our recommendation that practitioners should treat model outputs as untrustworthy signals requiring an independent validation layer. 
Furthermore, the documentation for Qwen-Agent explicitly advises using specific parsers to handle reasoning content failures~\cite{Qwen-Agent}, validating our classification of structural parsing errors as a distinct and critical class of semantic failure.

\textbf{Validation of Evolutionary Failure Patterns(Finding~\ref{fnd:evolution}).}
We observed that reliability challenges shift from foundational stability to subtle edge cases as models mature. This trajectory is mirrored by the release of DeepSeek-V3.1-Terminus in late 2025~\cite{DeepSeek-V3.1-Terminus}. 
Unlike typical performance-driven updates, this ``engineering-focused'' release specifically targeted community-reported stability issues—such as spurious language mixing—and introduced longitudinal tracking of known precision limitations. This industrial practice directly validates our finding that reliability is a dynamic lifecycle-oriented property that requires continuous version-specific documentation updates.

\section{Related Work}

\textbf{Empirical Study on DL System Failure.}
There is an extensive body of research characterizing failures in traditional DL systems. These studies have analyzed a broad range of issues, including general bug characteristics across various frameworks~\cite{TensorFlow2018ISSTA,DLFailure2019FSE,DLFailure2020ICSE,TensorFlow2020DASFAA,Junjie2023Tosem,DLFailure2019ASE,yianyi2019issre}, as well as specific fault classes like dependency bugs~\cite{DependencyBug2023FSE}, performance bottlenecks~\cite{fudan2022fse}, and API misuses~\cite{Chengcheng2021ICSE,DLAPI2024ICSE}. However, the paradigm shift from a ``build-from-scratch'' to a ``consume-and-integrate'' workflow in LLMs limits the applicability of this previous work. The dominant user challenges are no longer in areas like bespoke model architecture design or manual feature engineering. Instead, the focus of user-reported failures has moved to the complex, error-prone tasks of deployment, configuration, and integration with a brittle software stack. This creates a new landscape of reliability challenges that these foundational studies do not capture.

\textbf{Empirical Study on LLM System Failure.}
Recent research on LLM reliability is growing, but has largely examined the ecosystem from perspectives other than that of the open-source end-user. 
Several studies~\cite{Shangtang2024NSDI,MegaScale2024NSDI,Meta2025HPCA,grattafiori2024llama3herdmodels,GPUFailure2025} analyze failures that occur in LLM training phase. While important, these failures are internal to the model provider and are invisible to the users. Yan et al.~\cite{yan2025empiricalstudyproductionincidents} analyze analyzing production incidents in large-scale commercial GenAI cloud services from a service operator perspective. This provides an insightful ``service-centric'' view, examining system outages from the perspective of a LLM provider. Other studies~\cite{RAGFailure,chengcheng2025ICSE} have focused on an application-specific perspective, such as failures in Retrieval Augmented Generation (RAG) systems. Chen et al.~\cite{chen2025empiricalstudyopenaiapi} conducted a comprehensive study of 2,874 Stack Overflow discussions related to OpenAI APIs. Their work characterized the challenges faced by developers consuming black-box APIs, highlighting issues such as the complexities of prompt engineering, token-based cost management, and the handling of non-deterministic outputs.

In contrast, our work targets a distinct and unexplored layer: the runtime reliability of white-box models deployed on user-managed infrastructure. Unlike API users, these practitioners must manage the entire stack from CUDA kernels to quantization formats. Consequently, they encounter unique classes of dynamic failures that are invisible to static analysis and API consumers.

\section{Conclusion}
In this paper, we present the first comprehensive empirical study on the failures encountered by users when fine-tuning and deploying open-source LLMs. Through a systematic analysis of \num real-world issues, we provide a novel user-centric perspective on LLM reliability that complements prior work focused on training or production environments. Our central finding is that the dominant struggle for users is not with intrinsic model flaws, but with the immense complexity and brittleness of the surrounding deployment and configuration ecosystem. This insight reframes the challenge of LLM reliability, highlighting an urgent need for the community to develop better tooling, standardized practices, and automated methods to manage these ecosystem-driven complexities. 



\section{Data Availability}
We have made our collected dataset of \num LLM failures and analysis results publicly available~\cite{artifact}.

\bibliographystyle{ACM-Reference-Format}
\bibliography{ref}

\end{document}